\documentclass[11pt,a4paper]{article}

\usepackage[a4paper,margin=26mm]{geometry}
\usepackage{graphicx}
\usepackage{booktabs}
\usepackage{longtable}
\usepackage{array}
\usepackage{tabularx}
\usepackage{calc}
\usepackage{enumitem}
\usepackage{amsmath,amssymb}
\usepackage{microtype}
\usepackage{xurl}
\usepackage{seqsplit}
\usepackage[hidelinks]{hyperref}
\usepackage[font=small,labelfont=bf]{caption}
\usepackage{etoolbox}
\usepackage{eso-pic}

\setkeys{Gin}{width=\linewidth,keepaspectratio}
\providecommand{\tightlist}{%
  \setlength{\itemsep}{0pt}\setlength{\parskip}{0pt}}

\hypersetup{
  pdftitle={SoulAuth: An Actor-native Identity Architecture and Rust Reference Implementation for Humans and Long-lived AI Actors},
  pdfauthor={KunYuan; Harold Wang; Echo Li; Egusi Gui; Kiki Hu; Lucas Luo; Magnus Hu},
  pdfsubject={Actor-native identity and authentication security for Humans and long-lived AIActors},
  pdfkeywords={Actor-native Identity, AIActor, identity authentication security, Philosophical Engineering, identity attribution, Human-AI collaboration, Architecture Conformance}
}

\title{\textbf{SoulAuth: An Actor-native Identity Architecture and Rust Reference Implementation for Humans and Long-lived AI Actors}}
\author{%
  KunYuan \quad Harold Wang \quad Echo Li \quad Egusi Gui\\[0.25em]
  Kiki Hu \quad Lucas Luo \quad Magnus Hu\\[0.8em]
  \normalsize TRANTOR LABS, Singapore\\
  \normalsize Correspondence: \href{mailto:KunYuan@trantorlabs.sg}{KunYuan@trantorlabs.sg}\\
  \normalsize Software artifact: \url{https://github.com/TrantorLabs/SoulAuth}\\
  \normalsize Software license: Apache-2.0
}
\date{Preprint, Version 1.0 --- 10 September 2026}

\begin{document}
\AddToShipoutPictureFG*{%
  \AtPageUpperLeft{%
    \raisebox{-22mm}[0pt][0pt]{\hspace{26mm}\includegraphics[width=0.18\textwidth]{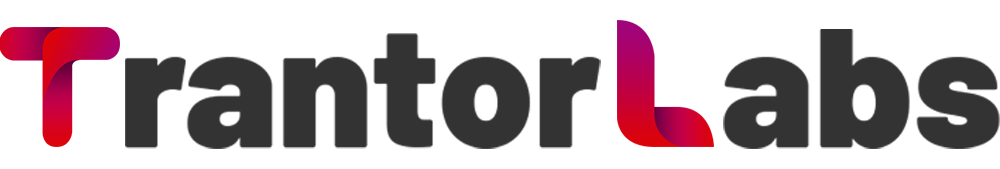}}%
  }%
}
\maketitle

\begin{abstract}
As some AI systems move from transient model invocations to long-running operation, Humans and long-lived AIActors are increasingly sharing the same digital infrastructure. Existing abstractions such as Account, Credential, Client, and Workload already address well-established problems within their respective domains. But when an AIActor must preserve identity continuity across changes in Credentials, Clients, AuthSessions, and runtimes, while historical authentication facts must continue to be attributed to the same subject, identity infrastructure must answer a more fundamental question: \textbf{where should the canonical continuity boundary be placed?}

This paper introduces \textbf{Actor-native Identity} and presents SoulAuth, an open-source reference implementation. Its central claim is that \textbf{any subject that must persist under its own identity and remain independently attributable should have an ActorIdentity that is not replaced by an Account, Credential, Client, AuthSession, or runtime instance.} SoulAuth therefore allows both Humans and long-lived AIActors to exist as first-class identity subjects, while HumanAccount, Credential, Client, IdentityBinding, AuthSession, protocol projections, and downstream Authority retain separate responsibilities and lifecycles. SoulAuth is an identity and authentication security architecture for Humans and long-lived AIActors. Its Rust reference implementation realizes this design as a runnable and inspectable software system, while Architecture Conformance links key architectural claims to runtime behavior, Schemas, machine-readable contracts, and verification evidence.

\textbf{Methodologically, the paper follows a Philosophical Engineering approach: it begins with conceptual clarification and structural analysis of subjecthood to determine what should serve as the bearer of persistent identity, then translates that judgment into an identity ontology, canonical invariants, lifecycle semantics, system responsibilities, engineering implementation, and inspectable conformance evidence.} In this way, the status of Humans and long-lived AIActors as identity subjects becomes more than a conceptual claim: it is translated into technical boundaries that an implementation can satisfy or violate and whose realization can be evaluated through evidence.

Existing work on machine identity, Workload Identity, DIDs, and agent identity already covers non-Human identities, persistent identifiers, Credentials, and parts of delegation. SoulAuth focuses on a structural choice above these mechanisms: \textbf{the canonical continuity boundary should be ActorIdentity, not Account, Credential, Client, Workload, AuthSession, Protocol Subject, or a particular runtime instance.} This choice gives systems in which Humans and long-lived AIActors coexist a stable subject for authentication and historical attribution, allowing downstream systems for accountability, delegation, authorization, and governance to refer to a well-defined ``who'' without equating identity-subject status with Authority or legal status.

The paper further evaluates Architecture Conformance on the fixed public SoulAuth \textbf{v0.1.0} release. The results show that the reference implementation realizes several core boundaries, including first-class status as identity subjects for Humans and long-lived AIActors, separation between Client and Actor, and separation between Authentication and Authority, while some implementation gaps remain in areas such as unified Credential modeling and historical attribution anchored to ActorIdentity. We therefore report the current reference implementation as exhibiting \textbf{partial conformance}, not full conformance.
\end{abstract}


\section{Introduction}\label{introduction}

As some AI systems move from one-shot model invocations to long-running operation and begin to persist across multiple AuthSessions, Clients, Credentials, and runtime lifecycles, a basic identity question becomes unavoidable: \textbf{when these peripheral objects keep changing, which object should identity infrastructure treat as the persistent subject?}

This is not merely an object-modeling problem inside an identity system. As AI systems participate over long periods in enterprise services, digital platforms, automated workflows, and Human-AI collaboration, authentication results increasingly become foundational inputs to attribution, authorization, governance, and audit. If infrastructure cannot reliably distinguish ``which persistent subject was authenticated,'' ``which software entity merely participated in the protocol,'' and ``which Credential was only proof material,'' then even a downstream system with complete authorization and governance mechanisms may still be built on the wrong subject relationships. Identity for long-lived AIActors is therefore both a computer systems problem and a sociotechnical infrastructure problem: \textbf{before a system can discuss what an AI may do, on whose behalf it acts, or what responsibility it should bear, it must first be able to answer reliably who it is.}

Modern digital identity systems already contain mature abstractions such as User, Account, Client, Service Account, and Workload, each designed for different settings including account management, protocol participation, and machine runtime identity. The problem is not that these abstractions have failed. The problem is that they answer different questions. Account describes an account. Credential describes how a subject proves itself. Client identifies software participating in a protocol. Workload describes a running computational workload. Treating any one of these objects as the default subject that persists across time gradually collapses responsibilities that were originally distinct.

We encountered this problem while building SoulAuth. Once native AIActor authentication entered the system, adding an AI user type or a new Credential could extend the authentication path, but it still did not answer the more fundamental question: \textbf{who, exactly, was authenticated?} We therefore chose not to continue extending the existing User model and instead restructured the subject model of the identity system. \texttt{ActorIdentity} became the canonical continuity boundary, while HumanAccount, Credential, Client, IdentityBinding, and AuthSession were separated from that boundary. HumanAccount thus returned to its proper role as a Human-specific account extension, while an AIActor no longer needed to depend on HumanAccount in order to possess a complete identity.

This restructuring is what we call \textbf{Actor-native Identity}. It is not a new login method. It is a structural choice about identity: the canonical continuity boundary is ActorIdentity, while Account, Credential, Client, IdentityBinding, AuthSession, protocol projections, and runtime each retain their own responsibilities and lifecycles. Humans and long-lived AIActors can therefore share the same higher-level identity model while continuing to use different Credentials, authentication methods, and lifecycles.

\textbf{What Humans and AIActors share is not the same authentication method, but a ``who'' that cannot be replaced by a peripheral object.} In SoulAuth, this is not a claim about the social status or legal rights of AI. It is an identity-security requirement: any subject that must authenticate independently and remain historically attributable needs a stable continuity boundary that cannot be replaced by a peripheral object.

We make this research path explicit as \textbf{Philosophical Engineering}. Rather than beginning with a new authentication component, we first clarify the distinct responsibilities of Account, Credential, Client, Workload, AuthSession, and runtime instance, and ask which object should serve as the persistent subject across those peripheral lifecycles. That conceptual and structural judgment is then expressed through ActorIdentity, relations in the identity ontology, canonical invariants, and lifecycle rules, and finally realized as system responsibilities, a Rust implementation, and Architecture Conformance evidence. Philosophy clarifies the foundational question of ``who''; engineering makes that answer operational; conformance evaluation checks whether the implementation remains faithful to the original subject definition.

This identity structure also requires a clear authentication boundary. SoulAuth identifies and authenticates an ActorIdentity, then conveys verified identity facts to downstream systems through AuthSession, Token, Claims, and related forms. Successful authentication does not itself create business permissions, governance decisions, or execution eligibility. In other words, \textbf{Authentication \ensuremath{\neq} Authority}: identity infrastructure answers ``who is the subject, and has that identity been proven?''; ``what may the subject do?'' remains the responsibility of downstream authorization, delegation, governance, or other domains.

\begin{figure}[htbp]
\centering
\includegraphics{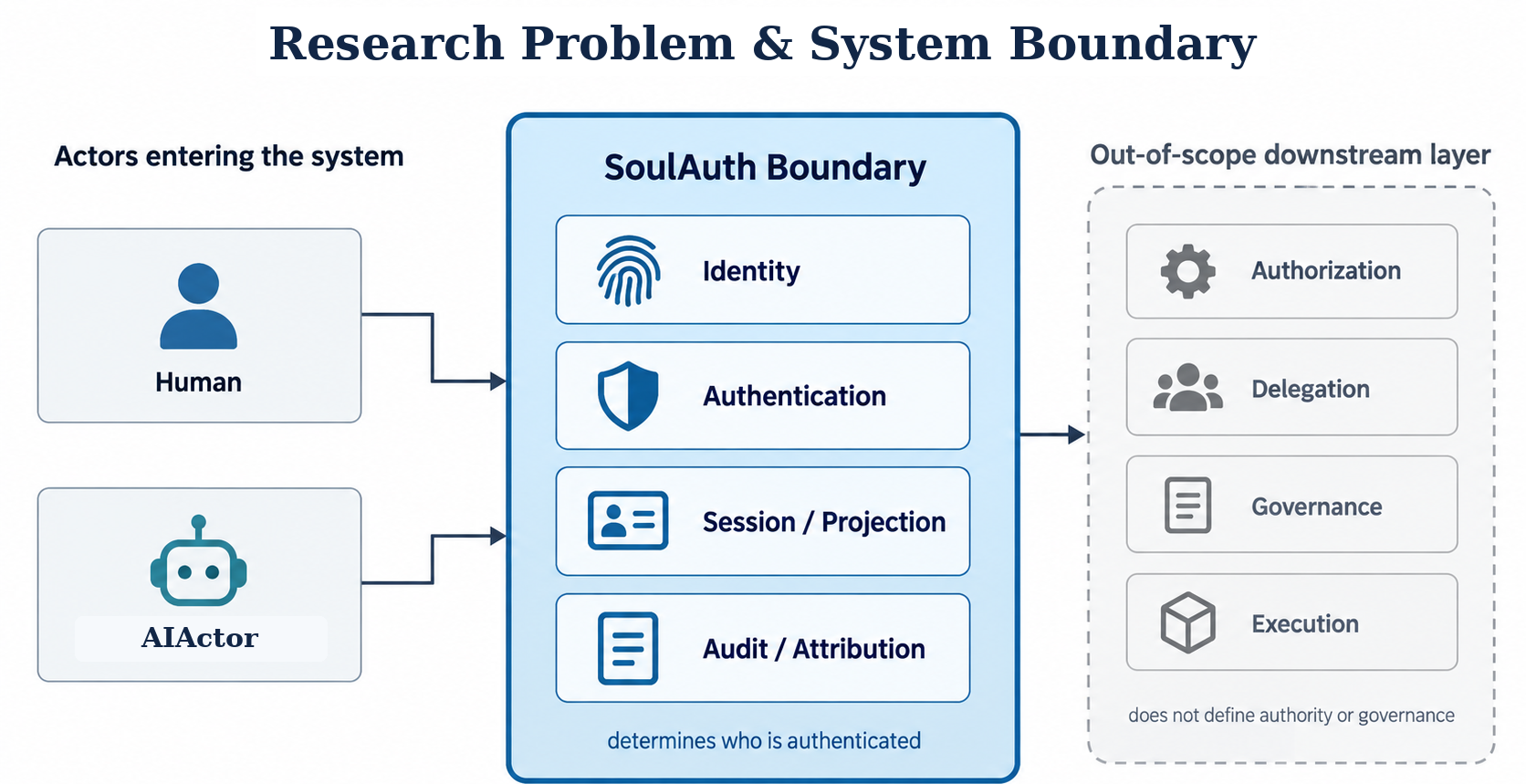}
\caption{Figure 1 \textbar{} SoulAuth Research Problem and System Boundary}
\end{figure}

Based on this model, we designed SoulAuth as Actor-native identity and authentication infrastructure for Humans and long-lived AIActors. The system organizes identity, authentication, AuthSession, tokens and federation, security, audit and attribution, control, and persistence around ActorIdentity as long-lived logical responsibilities, while separating those responsibilities from any specific process, container, or deployment topology. SoulAuth can run independently, or provide trusted authentication facts to other systems through explicit integration contracts.

This paper also provides an open-source \textbf{Rust reference implementation} of SoulAuth, turning the core Actor-native Identity objects, authentication paths, and responsibility boundaries into a runnable software system. We chose Rust not because a programming language can prove an architecture correct, but because it is an engineering vehicle well suited to security-sensitive identity infrastructure. The architecture defines objects, responsibilities, and boundaries; the Rust implementation makes those decisions compilable, deployable, and inspectable in a real system.

To keep the architecture from remaining only in design documents, SoulAuth also establishes an Architecture Conformance mechanism that links selected architectural claims to machine-readable contracts, runtime behavior, and verification evidence. This paper makes four main contributions:

\begin{enumerate}
\def\labelenumi{\arabic{enumi}.}
\tightlist
\item
  \textbf{Philosophical Engineering for conceptual clarification and subject modeling.} We begin from the question of what should count as the persistent subject that is identified, authenticated, and attributed. We clarify the distinct roles of Account, Credential, Client, Workload, AuthSession, and runtime instance, and establish \texttt{ActorIdentity} as the canonical continuity boundary, allowing both Humans and long-lived AIActors to become first-class identity subjects.
\item
  \textbf{Actor-native Identity Security Architecture.} We translate that subject-level judgment into object boundaries, the I1--I10 Canonical Invariants, lifecycle semantics, and a Logical Responsibility Architecture that separates identity, authentication, Authority, protocol projection, and historical evidence while preventing peripheral objects from regaining subject-defining authority.
\item
  \textbf{Identity continuity and identity-misattribution security analysis.} We distinguish identity continuity from authentication trust continuity and identify ``successful authentication attributed to the wrong persistent subject'' as a core security failure, with representative misattribution modes involving Credentials, Clients, runtime instances, Bindings, and Authority boundaries.
\item
  \textbf{Rust reference implementation and inspectable engineering evidence.} We provide an open-source Rust reference implementation and use Architecture Conformance, negative testing, a fixed source revision, and CI evidence to connect normative judgments about identity subjects to observable engineering states such as PASS, PARTIAL, and FAIL.
\end{enumerate}

This paper does not address whether AI systems are conscious or possess legal personhood, nor does it attempt to merge identity, authorization, governance, and execution into a single system. Its scope is limited to the identity and authentication layer: when Humans and long-lived AIActors share identity infrastructure, the canonical continuity boundary represented by ActorIdentity must remain intact across authentication, runtime evolution, and historical attribution. SoulAuth starts from a simple design principle: first determine who the Actor is, then determine how that Actor proves itself.

\section{Problem Model, Design Requirements, and Boundaries}\label{problem-model-design-requirements-and-boundaries}

Actor-native Identity does not begin from the premise that existing identity systems cannot support machines. HumanAccount, OAuth Client, Service Account, Workload Identity, and a wide range of Credential mechanisms already serve many mature use cases. What must be reconsidered is a different issue: when a non-Human computational system must preserve its own identity across repeated authentication and changing runtime environments, and must remain a persistent subject of authentication and attribution, objects with different original responsibilities can no longer be collapsed into a single notion of subjecthood.

\subsection{Misaligned Identity Objects}\label{misaligned-identity-objects}

Identity systems commonly contain multiple objects that carry identifiers. HumanAccount describes how a Human manages an account. Credential describes the material through which a subject proves identity. Client identifies software participating in OAuth or OIDC. Workload describes a running computational workload. Service Account is commonly used to allow non-Human software to access protected resources. All of these objects may participate in authentication or authorization, but they answer different questions.

In most Human-centered applications, these distinctions are not always salient. A Human often uses a system through a limited set of Accounts and Credentials, and the account, login subject, and ultimately authenticated subject remain highly correlated over time. As a result, an implementation may fail to separate these semantics rigorously without immediately exposing the problem.

Long-running AI systems change that condition. When Credential, Client, session, and runtime can all change, promoting any one of them directly into the persistent identity subject forces that object to carry semantics beyond its original responsibility. An OAuth Client, Service Account, Credential, or Workload may be related to an AIActor, but none of them should determine who that Actor is merely by existing.

\subsection{Minimal Engineering Definition and Identity Requirements of an AIActor}\label{minimal-engineering-definition-and-identity-requirements-of-an-aiactor}

In this paper, \texttt{AIActor} is an engineering category at the identity layer. It makes no claim about consciousness, personhood, or legal status.

When a non-Human computational entity must authenticate under its own identity, remain recognizable as the same subject after legitimate changes to Credential, session, Client, or runtime instance, and retain stable attribution of historical authentication activity, identity infrastructure needs to provide that entity with an independent representation of persistent identity.

Not every Bot, Model Invocation, Agent, or background process must therefore become an AIActor. A computation that exists only for a single request and whose responsibility is fully carried by an upstream application identity may not need its own identity. By contrast, an AI system that must persist over time, authenticate independently, rotate Credentials, and maintain its own authentication history creates a different identity requirement.

This requirement means that the existence of a HumanAccount cannot be a prerequisite for an AIActor to possess a complete identity. Humans and AIActors may use different Credentials and authentication methods, but both must be independently recognizable by the identity system as a ``who.'' What is unified is their status as identity subjects, not their authentication mechanism or implementation structure.

\subsection{Identity Continuity and Historical Attribution}\label{identity-continuity-and-historical-attribution}

Persistent identity first requires the system to continue recognizing the same Actor across legitimate changes to peripheral objects. For the same Actor, Credentials may rotate, Clients may change, AuthSessions may end and be re-established, and runtimes may be updated. These changes affect how the Actor proves itself at a given moment and what authentication state currently holds, but they must not create a new ActorIdentity merely by changing.

Identity continuity therefore does not mean that all data are immutable forever. Profile, Credential, Binding, and other peripheral state can each have their own lifecycle. What must remain stable is the system's ability to determine whether it is still dealing with the same Actor after those objects change legitimately.

The same requirement also runs backward through time. A system must not only determine whether an Actor remains the same after a change; it must also ensure that current state changes do not reinterpret past authentication facts. If historical attribution depends on the current Profile, current IdentityBinding, or another mutable state object, then changing that state can change the apparent subject of past events.

SoulAuth therefore requires historical authentication and audit events to remain associated with the stable ActorIdentity to which they were attributed when the events occurred. This design requirement can be understood as continuity in two directions: forward, peripheral state changes must not silently replace the Actor; backward, current state changes must not rewrite identity attribution for events that have already happened.

\subsection{Cross-Domain Identity Relationships Must Be Explicit}\label{cross-domain-identity-relationships-must-be-explicit}

An Actor may appear in more than one identity domain. An External IdP, an organization-internal identity system, a SoulAuth ActorIdentity, and other external identity systems may each have their own Subjects and identifiers. The presence of identical or similar identifiers in two domains is not enough to prove that they naturally refer to the same identity.

SoulAuth uses \texttt{IdentityBinding} to represent an explicit cross-domain relationship. A Binding declares and maintains a constrained correspondence; it does not merge two identity domains into a single Source of Truth. Therefore:

\begin{quote}
\textbf{Relationship \ensuremath{\neq} Identity Equivalence.}
\end{quote}

Similarity in email, display name, or external Subject string cannot itself establish identity equivalence, and the existence of an IdentityBinding cannot replace authentication. Cross-domain systems may establish relationships, but each identity domain retains ownership of its own identity facts and definitional authority within that domain.

\subsection{Authentication Boundary and Non-Goals}\label{authentication-boundary-and-non-goals}

Actor-native Identity solves the problem of ``who,'' not the entire problem of ``what may this subject do.'' SoulAuth therefore keeps identity, authentication, and Authority as three distinct layers: identity defines the subject being identified; authentication determines whether current evidence is sufficient to prove that identity; Authority determines what powers an identified and authenticated subject may exercise in a particular context.

These layers depend on one another, but they must not be compressed into a single object:

\begin{quote}
\textbf{Identity \ensuremath{\neq} Authentication \ensuremath{\neq} Authority.}
\end{quote}

A successful authentication establishes only those authentication facts that the applicable authentication contract permits it to express. It does not automatically create application permissions, governance Authority, or execution Authority. SoulAuth can reliably tell a relying system ``which authenticated Actor this request comes from,'' but it does not infer what that Actor may therefore do.

This boundary also defines the paper's non-goals. AIActor is only an identity-layer engineering category. SoulAuth does not determine whether an AI is conscious or a legal person, nor does it attempt to recombine identity, Authority, and execution into one system.

Taken together, these requirements mean that SoulAuth does not treat ``uses AI'' as a sufficient condition for creating an AIActor identity. \textbf{Only when a non-Human computational entity must persist under its own identity across repeated authentication, Credentials, Clients, AuthSessions, or runtime instances, and become an independent subject of historical attribution, does identity infrastructure need to establish a distinct ActorIdentity for it.} The common ground between Humans and AIActors is therefore not that they authenticate in the same way, but that each has a persistent subject boundary that cannot be replaced by a peripheral object.

This first-class subject status exists only at the identity layer. It does not automatically imply identical Credentials, authentication methods, Authority, capabilities, responsibilities, or legal status.

\section{Actor-native Identity Model}\label{actor-native-identity-model}

\subsection{ActorIdentity: The Canonical Continuity Boundary}\label{actoridentity-the-canonical-continuity-boundary}

SoulAuth defines \texttt{ActorIdentity} as the \textbf{canonical continuity boundary} within the identity domain. It answers the most basic question in an identity system: \textbf{to which subject do these Credentials, authentication facts, AuthSessions, IdentityBindings, and attributions ultimately belong?}

ActorIdentity is not simply a traditional \texttt{User} object under a new name. An ActorIdentity may have different relationships with HumanAccount, Credential, IdentityBinding, AuthSession, and Client, but none of those objects can substitute for ActorIdentity itself. ActorIdentity preserves the continuity of ``who''; peripheral objects separately carry responsibilities such as account management, proof of identity, protocol participation, cross-domain relationships, and authentication state.

ActorIdentity is also not inherently a globally universal identifier across all systems. It is first canonical within the SoulAuth identity domain. An external Subject, protocol-level Subject, or identity object in another domain may be related to a SoulAuth ActorIdentity through an explicit mapping, projection, or IdentityBinding, but similarity in identifier, email, or other representation cannot make the two identities automatically equivalent.

The value of ActorIdentity is therefore not that it adds another ID. Its value is that it establishes a stable identity boundary. Credentials may rotate around it, Clients may change, AuthSessions may begin and end, and IdentityBindings may be created or revoked, but these peripheral changes do not acquire the power to redefine the Actor.

\subsection{ActorKind and First-Class Identity-Subject Status}\label{actorkind-and-first-class-identity-subject-status}

SoulAuth's canonical \texttt{ActorKind} includes \texttt{Human} and \texttt{AIActor}. ActorKind belongs to the identity core of ActorIdentity; it is not a display property that can be changed through ordinary Profile updates.

Humans and AIActors have equal first-class status as identity subjects. ``First-class'' has a precise engineering meaning here: each can independently exist within the SoulAuth identity domain as a subject that can be identified, authenticated, and attributed, and each can establish authentication relationships around its own ActorIdentity.

This parity applies only at the level of identity-subject status. It does not require Humans and AIActors to have the same Credentials, authentication methods, lifecycles, or Authority. Actor-native Identity unifies the higher-level identity model, not the concrete implementation. Humans may retain Human-specific accounts and authentication methods; AIActors may use Credentials suited to their mode of operation. Neither must imitate the other in order to participate in the same identity semantics.

\begin{figure}[htbp]
\centering
\includegraphics{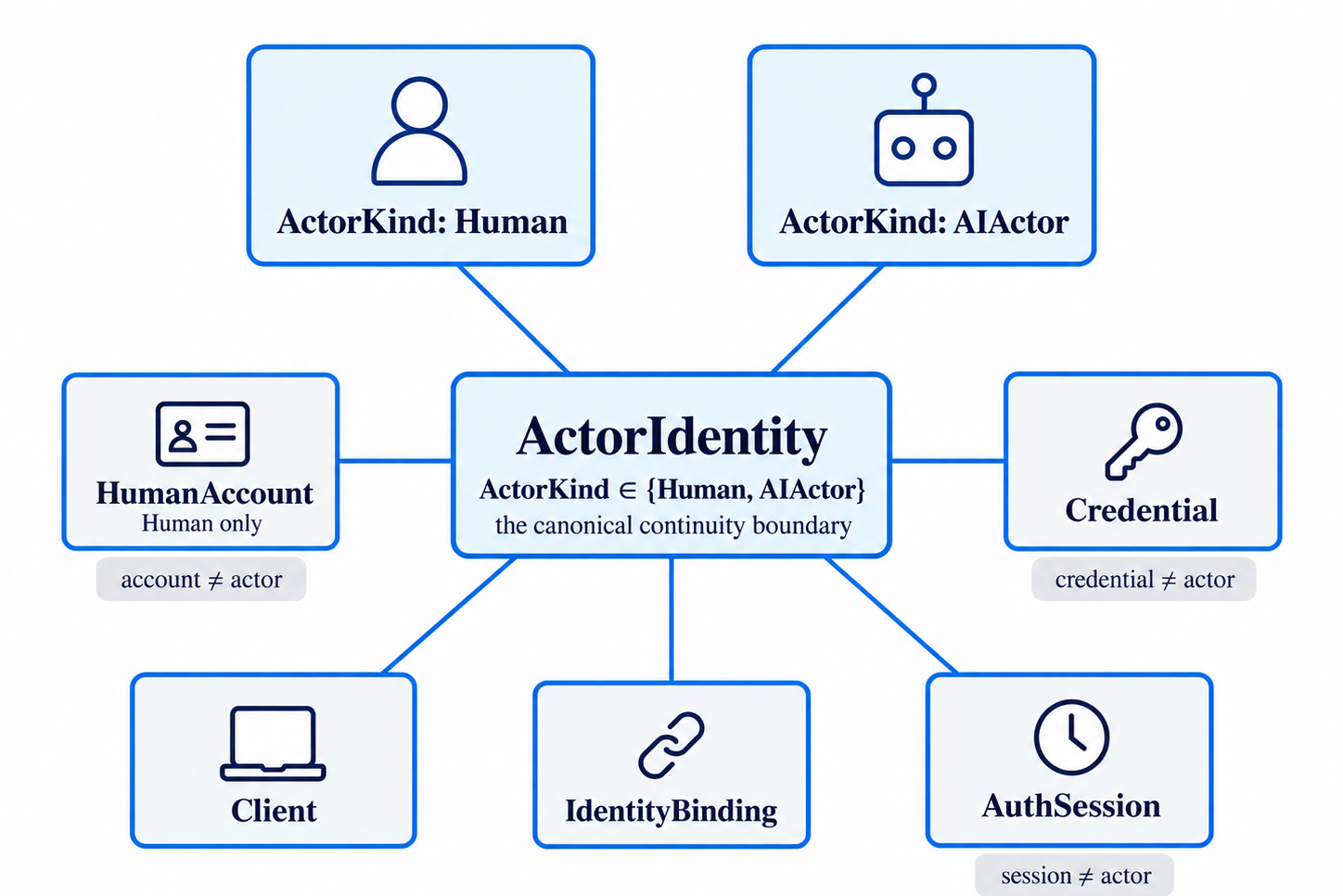}
\caption{Figure 2 \textbar{} Actor-native Identity Model}
\end{figure}

\subsection{Identity Objects Around ActorIdentity}\label{identity-objects-around-actoridentity}

Once ActorIdentity becomes the canonical continuity boundary, HumanAccount, Credential, Client, IdentityBinding, and AuthSession continue to perform necessary functions, but none retains the authority to define ActorIdentity.

\textbf{HumanAccount} is a Human-specific account extension associated with a Human ActorIdentity. It may carry email, username, and other account-related state, but those attributes do not themselves define who the Actor is. Changes to HumanAccount should not create a new ActorIdentity; an AIActor may have a complete identity even without any HumanAccount.

\textbf{Credential} represents the material or authentication capability through which an Actor proves itself. It answers ``how is the Actor proving itself?'' rather than ``who is the Actor?'' A single ActorIdentity may add, rotate, or revoke Credentials without thereby creating a new ActorIdentity.

\textbf{Client} represents the software entity participating in an identity protocol. It answers ``which software is interacting with the identity system?'' rather than ``which Actor is being authenticated?'' The same Actor may use SoulAuth through different Clients, and the same Client may serve different Actors. A Client's protocol identity therefore cannot be automatically elevated into the authenticated subject.

\textbf{IdentityBinding} represents an explicit relationship between identity domains. It can establish a constrained correspondence between a SoulAuth ActorIdentity and an external identity subject, but the Binding is not itself an Actor, nor does it merge the two identity domains into one ontology.

\textbf{AuthSession} represents continuity of authentication over a bounded period. It can carry the fact that an Actor has authenticated and provide context for later Tokens, Claims, or other identity projections, but it remains part of the authentication domain rather than ActorIdentity itself.

\subsection{Object Relationships and Lifecycles}\label{object-relationships-and-lifecycles}

The point of the Actor-native Identity model is not to maximize the number of objects, but to ensure that each object carries only its own responsibility and has its own lifecycle.

\begin{longtable}[]{@{}
  >{\raggedright\arraybackslash}p{(\columnwidth - 6\tabcolsep) * \real{0.2500}}
  >{\raggedright\arraybackslash}p{(\columnwidth - 6\tabcolsep) * \real{0.2500}}
  >{\raggedright\arraybackslash}p{(\columnwidth - 6\tabcolsep) * \real{0.2500}}
  >{\raggedright\arraybackslash}p{(\columnwidth - 6\tabcolsep) * \real{0.2500}}@{}}
\toprule\noalign{}
\begin{minipage}[b]{\linewidth}\raggedright
Object
\end{minipage} & \begin{minipage}[b]{\linewidth}\raggedright
Core Responsibility
\end{minipage} & \begin{minipage}[b]{\linewidth}\raggedright
Lifecycle
\end{minipage} & \begin{minipage}[b]{\linewidth}\raggedright
Key Boundary
\end{minipage} \\
\midrule\noalign{}
\endhead
\bottomrule\noalign{}
\endlastfoot
\texttt{ActorIdentity} & Canonical continuity boundary & Actor lifecycle & \ensuremath{\neq} HumanAccount, Credential, Client, IdentityBinding, AuthSession \\
\texttt{HumanAccount} & Human-specific account & Account lifecycle & \ensuremath{\neq} ActorIdentity \\
\texttt{Credential} & Capability for proving identity & Credential lifecycle & \ensuremath{\neq} ActorIdentity \\
\texttt{Client} & Protocol software identity & Client lifecycle & \ensuremath{\neq} ActorIdentity \\
\texttt{IdentityBinding} & Cross-domain identity relationship & Binding lifecycle & \ensuremath{\neq} ActorIdentity; does not merge the Sources of Truth of either side \\
\texttt{AuthSession} & Bounded authentication continuity & AuthSession lifecycle & \ensuremath{\neq} ActorIdentity \\
\end{longtable}

The key property of this structure is lifecycle independence. A normal change to one object must not silently alter the identity semantics of another through an implicit relationship. Credential Rotation is not Actor Replacement. Binding Revocation is not Actor Retirement. AuthSession Revocation is not Credential Revocation. Client Disable is not Actor Suspension.

This separation lets the system support both change and continuity. An Actor does not remain the same because all peripheral objects are frozen forever. It remains the same because, after those objects change according to their own rules, the system can still determine reliably whether it is dealing with the same Actor.

Actor Retirement likewise does not mean that a previously used ActorIdentity identifier may be reassigned. If a retired ActorIdentity identifier were reassigned to another Actor, historical attribution would lose a unique interpretation. Retirement therefore terminates future valid use; it does not erase the historical uniqueness of the ActorIdentity identifier.

\subsection{Canonical Invariants}\label{canonical-invariants}

The preceding model can be compressed into a set of Canonical Invariants that do not depend on a particular database, protocol, or deployment model. They define the minimum semantic boundary of Actor-native Identity, not one specific implementation.

\begin{enumerate}
\def\labelenumi{\arabic{enumi}.}
\tightlist
\item
  \textbf{I1 \textbar{} ActorIdentity \ensuremath{\neq} HumanAccount.}
\item
  \textbf{I2 \textbar{} ActorIdentity \ensuremath{\neq} Credential.}
\item
  \textbf{I3 \textbar{} ActorIdentity \ensuremath{\neq} Client.}
\item
  \textbf{I4 \textbar{} ActorIdentity \ensuremath{\neq} IdentityBinding.}
\item
  \textbf{I5 \textbar{} Credential Rotation \ensuremath{\neq} Actor Replacement.}
\item
  \textbf{I6 \textbar{} Client Change \ensuremath{\neq} Actor Replacement.}
\item
  \textbf{I7 \textbar{} Humans and AIActors have equal first-class status as identity subjects; AIActor status as an identity subject does not depend on HumanAccount, nor does it require the same Credentials, authentication methods, lifecycles, or Authority as a Human.}
\item
  \textbf{I8 \textbar{} Authentication \ensuremath{\neq} Authority.}
\item
  \textbf{I9 \textbar{} Current mutable identity state must not redefine historical attribution.}
\item
  \textbf{I10 \textbar{} A retired ActorIdentity identifier must not be reassigned to another Actor.}
\end{enumerate}

Together, these ten invariants form the minimum semantic skeleton of the SoulAuth Actor-native Identity model. They do not prescribe a particular database, Credential, authentication protocol, or deployment topology. They specify a more fundamental structural relation: \textbf{ActorIdentity is the canonical continuity boundary; other objects operate around it, but no peripheral object may redefine who the Actor is merely because its own state or lifecycle changes.}

More generally, \textbf{AuthSession, Workload, Protocol Subject, protocol projection, and runtime instance may reference, carry, or project ActorIdentity, but none may become the canonical continuity boundary in the opposite direction.}

\section{SoulAuth System Architecture}\label{soulauth-system-architecture}

\subsection{Organizing Identity Infrastructure by Logical Responsibility}\label{organizing-identity-infrastructure-by-logical-responsibility}

On top of the Actor-native Identity model, SoulAuth organizes the system around durable logical responsibilities rather than a particular deployment model, database structure, or source-code layout. The principal responsibility domains are as follows.

\begin{longtable}[]{@{}
  >{\raggedright\arraybackslash}p{(\columnwidth - 2\tabcolsep) * \real{0.5000}}
  >{\raggedright\arraybackslash}p{(\columnwidth - 2\tabcolsep) * \real{0.5000}}@{}}
\toprule\noalign{}
\begin{minipage}[b]{\linewidth}\raggedright
Architectural Responsibility
\end{minipage} & \begin{minipage}[b]{\linewidth}\raggedright
Core Responsibility
\end{minipage} \\
\midrule\noalign{}
\endhead
\bottomrule\noalign{}
\endlastfoot
\textbf{Clients} & Represent software participants interacting with SoulAuth \\
\textbf{Access and Protocol Boundary} & Receive and normalize supported identity-protocol requests \\
\textbf{Identity Domain} & Maintain ActorIdentity and identity relationships \\
\textbf{Authentication Core} & Determine whether current authentication evidence is valid \\
\textbf{AuthSession} & Maintain bounded authentication state \\
\textbf{Tokens and Federation} & Express authenticated identity facts outward through protocols \\
\textbf{Control Plane} & Manage SoulAuth itself \\
\textbf{Security Protections} & Protect identity and authentication lifecycles across domains \\
\textbf{Audit and Attribution} & Record identity and authentication facts and their attribution \\
\textbf{Persistence and Infrastructure} & Persist domain state and connect to external infrastructure \\
\end{longtable}

\begin{figure}[htbp]
\centering
\includegraphics{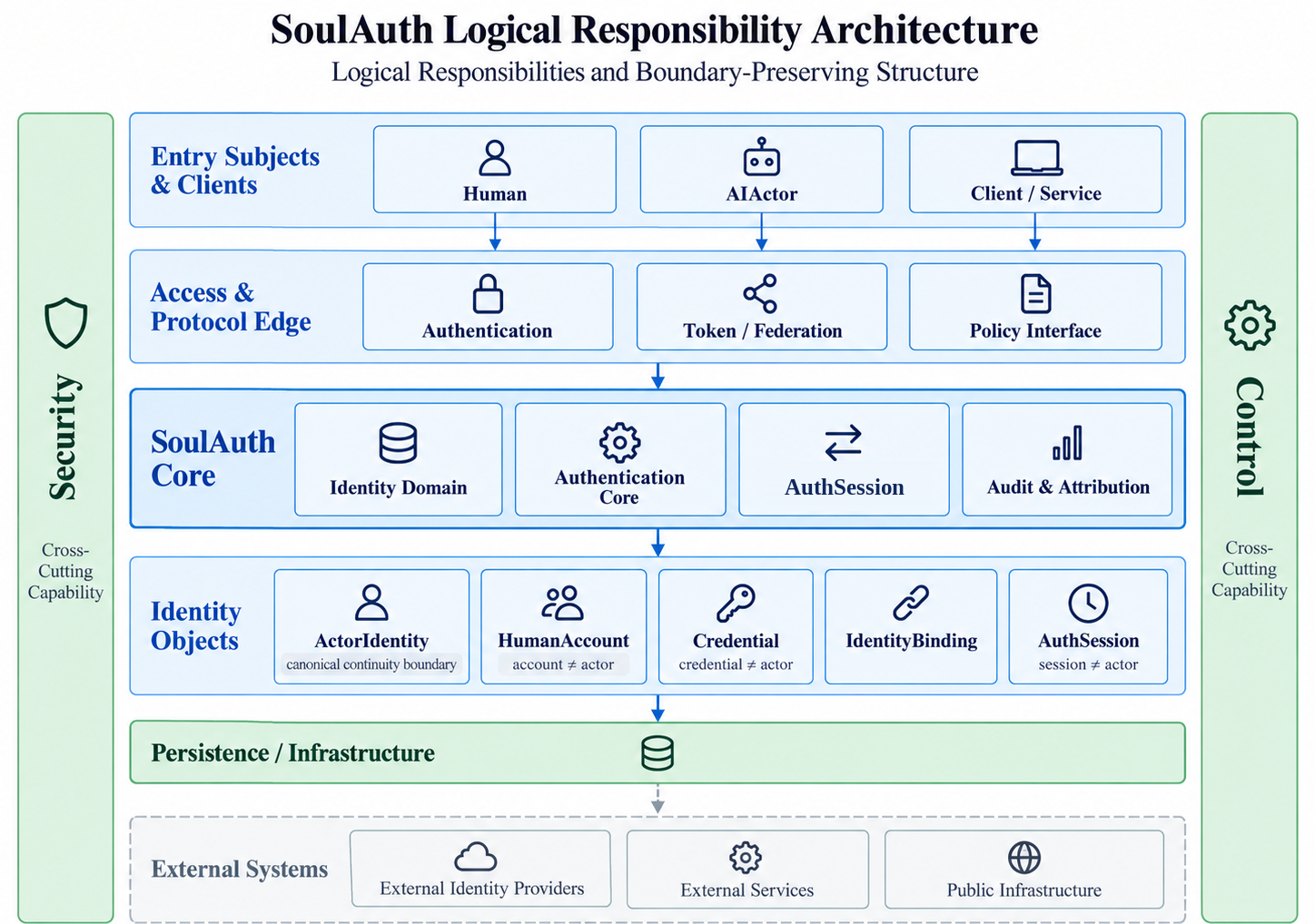}
\caption{Figure 3 \textbar{} SoulAuth Logical Responsibility Architecture}
\end{figure}

This is a \textbf{Logical Responsibility Architecture}, not a fixed runtime pipeline. Different authentication, federation, or administrative requests may follow different runtime paths as long as they preserve the same responsibility boundaries.

Logical responsibilities are likewise independent of deployment units. A single Rust process may host several architectural responsibilities, and one responsibility may later be split across multiple runtime units. An architectural component is therefore not the same thing as a process, container, or microservice. SoulAuth fixes responsibilities and boundaries, not one deployment topology.

The architecture follows one common rule: \textbf{each responsibility domain defines only the facts it owns; data relationships or implementation convenience do not confer the right to define another domain's semantics.}

\subsection{Identity Domain and Authentication Core}\label{identity-domain-and-authentication-core}

The center of SoulAuth consists of two closely collaborating responsibilities that must nevertheless remain distinct: the \textbf{Identity Domain} and the \textbf{Authentication Core}.

The Identity Domain answers ``who.'' It maintains ActorIdentity and related identity relationships so that both Humans and AIActors can be identified reliably as independent subjects. The Authentication Core answers ``has this identity been proven?'' It consumes Credentials or other authentication evidence and determines, under the current authentication contract, whether the proof is sufficient.

Therefore:

\begin{quote}
\textbf{Identity Domain \ensuremath{\neq} Authentication Core.}
\end{quote}

The existence of an ActorIdentity does not imply that authentication will succeed, nor can a successful authentication redefine ActorIdentity on the basis of a Credential or proof result. Humans and AIActors may use different Credentials and authentication methods without requiring two separate identity systems. Every authentication path must ultimately resolve to a specific ActorIdentity and produce an authentication result belonging to that Actor.

\subsection{Authentication State and Protocol Projection}\label{authentication-state-and-protocol-projection}

After authentication completes, SoulAuth must handle two separate questions: how an established authentication fact persists for a bounded period, and how that fact is conveyed to external relying systems.

\texttt{AuthSession} handles the first problem. It represents continuity of an established authentication fact over a bounded period. It may be created, expire, or be revoked without changing ActorIdentity itself.

\textbf{Tokens and Federation} handle the second problem. OIDC, Token, Claims, UserInfo, and other protocol forms project identity and authentication facts already established by SoulAuth into representations that relying systems can understand and verify. The critical boundary is:

\begin{quote}
\textbf{Projection \ensuremath{\neq} Source of Truth.}
\end{quote}

Tokens, Claims, and protocol-level Subjects may be relied upon within the declared trust contract, but they remain representations of upstream identity facts rather than ActorIdentity itself.

Federation follows the same principle. An external identity provider may supply authentication facts or an external Subject, but the relationship between identity domains must still be established explicitly. A protocol exchanges and conveys identity facts; it does not thereby merge the Sources of Truth of the participating identity domains.

\subsection{Control, Security, and Audit}\label{control-security-and-audit}

The \textbf{Control Plane}, \textbf{Security Protections}, and \textbf{Audit and Attribution} span the entire identity lifecycle rather than being appended only after a login completes.

The \textbf{Control Plane} manages SoulAuth itself, including controlled administration of identities, Credentials, Clients, and relevant infrastructure state. It carries SoulAuth's internal management responsibilities, not the governance or execution Authority of downstream systems.

\textbf{Security Protections} span identity, authentication, AuthSession, Token, Client, and infrastructure boundaries, preserving security constraints and trust boundaries across these responsibilities. Security is not an add-on outside the architecture; it is a cross-cutting responsibility throughout the identity and authentication lifecycle.

\textbf{Audit and Attribution} preserve facts about identity and authentication events that have already occurred and the Actor to which those facts should be attributed. Audit is therefore not a copy of current identity state; it is a form of historical evidence.

From this perspective, SoulAuth distinguishes three different kinds of facts:

\begin{itemize}
\tightlist
\item
  \textbf{Canonical domain state}: current identity facts maintained by the Identity Domain;
\item
  \textbf{Trusted projection}: trusted representations for relying parties, such as Tokens and Claims;
\item
  \textbf{Historical evidence}: records of events that occurred in the past, such as audit events.
\end{itemize}

All three may be related through ActorIdentity, but none substitutes for another:

\begin{quote}
\textbf{Domain State \ensuremath{\neq} Projection \ensuremath{\neq} Historical Evidence.}
\end{quote}

\begin{figure}[htbp]
\centering
\includegraphics{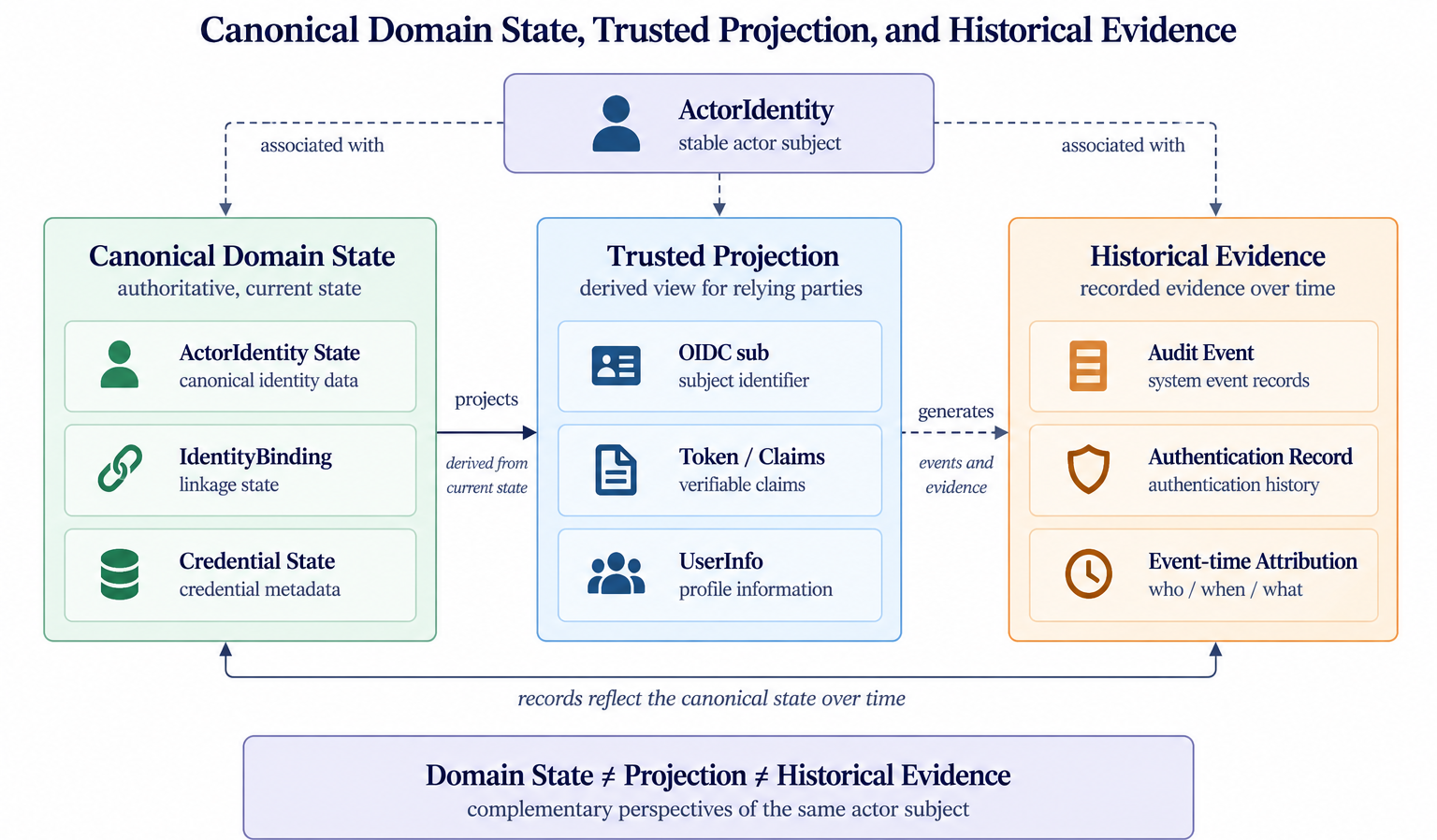}
\caption{Figure 4 \textbar{} Canonical Domain State, Trusted Projection, and Historical Evidence}
\end{figure}

This separation lets current identity state, external protocol representation, and historical attribution evolve independently without allowing a change in one layer to redefine the other two.

\subsection{Persistence and Infrastructure}\label{persistence-and-infrastructure}

SoulAuth's identity, authentication, session, and audit state must ultimately be carried by concrete infrastructure, but infrastructure does not own the definition of upstream identity semantics.

Persistence and infrastructure sit at the supporting layer of the architecture. Identity, Credential, AuthSession, OIDC, security, and audit may each have their own logical stores, but logical storage is not the same thing as a physical database. Therefore:

\begin{quote}
\textbf{One Database \ensuremath{\neq} One Domain.}
\end{quote}

Multiple logical domains may share a physical database, and a single logical domain may span multiple storage implementations. Database topology does not erase logical boundaries.

Likewise:

\begin{quote}
\textbf{Persistence Schema \ensuremath{\neq} Canonical Ontology.}
\end{quote}

Domain objects may be realized through different persistence structures, and a table may be only an implementation detail. A concrete Schema carries domain state; it does not gain the right to redefine ActorIdentity, Credential, or lifecycle semantics for reasons of storage convenience.

Adapters and external infrastructure follow the same rule. Persistence adapters, External IdPs, key managers, and similar systems provide capabilities to SoulAuth through explicit contracts; they do not become owners of the domain ontology. External consumers should likewise obtain identity and authentication facts through supported integration contracts rather than depending on SoulAuth's private persistence structures.

SoulAuth can therefore allow its database, code organization, runtime units, and deployment topology to evolve while preserving more stable architectural relations: the Identity Domain maintains ``who''; authentication determines ``has this identity been proven?''; AuthSession and federation extend and project authentication facts; control, security, and audit span the lifecycle; and persistence and infrastructure carry those semantics without redefining them.

\section{Identity Continuity and Authentication Semantics}\label{identity-continuity-and-authentication-semantics}

\subsection{Authentication Runtime Model}\label{authentication-runtime-model}

ActorIdentity solves the problem of ``who,'' but the existence of an ActorIdentity does not mean that the evidence in the current request has proven that identity. Authentication does not create or redefine the Actor. Its responsibility is to determine, within an explicit identity context, whether the proof currently presented is sufficient to establish an authentication fact about that ActorIdentity.

SoulAuth abstracts an authentication event as the following runtime relation:

\begin{quote}
\textbf{Identity Context + Authentication Source → Evidence or Assertion → Verification → Authentication Result → Optional AuthSession → Protocol or Integration Projection}
\end{quote}

Authentication Source is an upper-level concept. A local Credential is one important source, but not a universal prerequisite for every authentication path. An external authentication source that has been verified under a supported federation contract can enter the same model.

This runtime relation requires several objects that implementations often collapse to remain distinct. \texttt{Credential} represents a long-lived authentication capability owned by an Actor. Authentication evidence is the proof actually submitted in a particular authentication event. Authentication Result is the fact established after verification. \texttt{AuthSession} represents whether an already established authentication fact may continue to be reused within a bounded lifecycle. These objects are related, but they operate on different timescales and carry different responsibilities.

A successful authentication therefore establishes the following: \textbf{an ActorIdentity already resolved in the Identity Domain has been proven under a particular method, time, context, and set of verification conditions.} It does not redefine ActorIdentity, create permanent trust, or automatically create downstream Authority.

\subsection{Authentication Paths for Humans and AIActors}\label{authentication-paths-for-humans-and-aiactors}

Actor-native Identity unifies the subject model, not the authentication method. Humans and AIActors may use different authentication sources, Credentials, and evidence, as long as the final authentication process verifies a clearly identified ActorIdentity and produces a semantically well-defined authentication result.

Human authentication may rely on passwords, MFA, supported federation mechanisms, or other methods. Long-lived Credentials must remain distinct from the evidence submitted in a particular authentication event. If an authentication contract requires multiple factors, the success of one factor cannot automatically mean that authentication as a whole has completed.

AIActors may use authentication mechanisms suited to machine subjects. The long-lived object is the authentication capability associated with ActorIdentity; the current request carries proof formed for a particular context. Different ActorKinds may therefore follow different authentication paths, and ActorKind itself is not equivalent to one authentication method.

What is unified is the upper-level semantics rather than the protocol flow. For both Humans and AIActors, the system must answer the same questions: which ActorIdentity is being authenticated, which authentication source is being used, what evidence has been submitted, under what conditions verification succeeds, and what authentication result is established.

Federation is a different dimension. It is not a third ActorKind; it is one possible Authentication Source. An External IdP may provide a protocol-verified authentication fact, but SoulAuth must still resolve the corresponding ActorIdentity in its own Identity Domain and establish local authentication semantics through an explicit IdentityBinding and trust contract. Subject type answers ``who''; Authentication Source answers ``where did the proof come from?'' The two are independent.

\begin{figure}[htbp]
\centering
\includegraphics{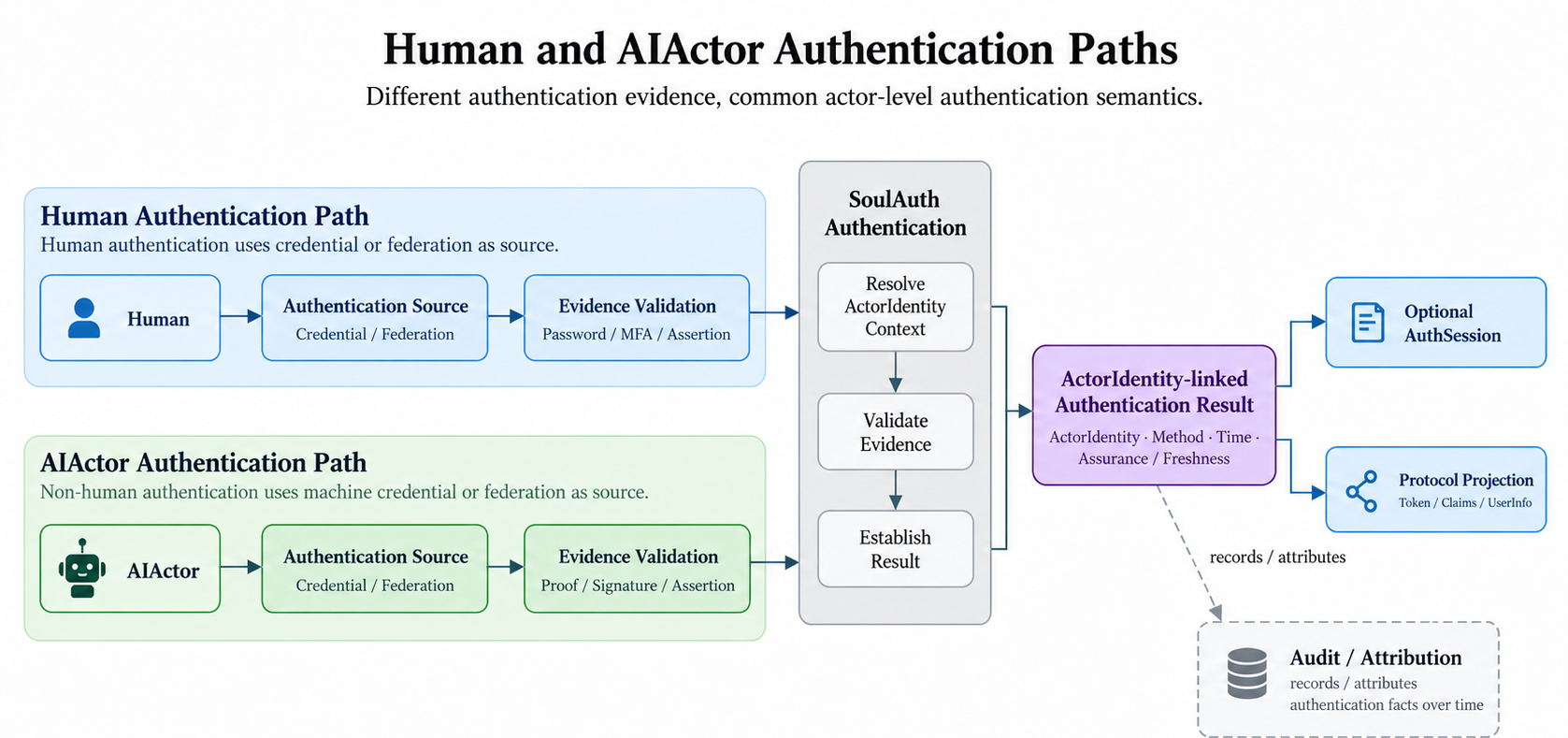}
\caption{Figure 5 \textbar{} Human and AIActor Authentication Paths}
\end{figure}

\subsection{Authentication Result, Freshness, and AuthSession}\label{authentication-result-freshness-and-authsession}

A successful authentication should express more than success or failure. To let downstream systems interpret the established authentication fact correctly, the Authentication Result also needs enough context to describe the conditions under which that fact holds, such as the authenticated Actor, ActorKind, authentication method, authentication time, applicable assurance level, and freshness information.

Assurance level and freshness must not be conflated. Assurance level describes the verification conditions under which an authentication fact was established; freshness describes how recently that proof was established relative to the current request. A higher authentication assurance level does not automatically grant greater Authority. Likewise, an AuthSession that has not expired does not, merely by virtue of its lifetime, prove that the authentication fact it carries is fresh enough for every new trust requirement.

SoulAuth therefore keeps the following relations distinct:

\begin{quote}
\textbf{Authentication Method \ensuremath{\neq} Authentication Assurance Level; Authentication Freshness \ensuremath{\neq} AuthSession Lifetime; Higher Authentication Assurance \ensuremath{\neq} Greater Authority.}
\end{quote}

The responsibility of \texttt{AuthSession} is to extend an established authentication fact within explicit boundaries. It is not ActorIdentity, not Credential, and not the Token itself. Whether an authentication event creates an AuthSession, and under what conditions that AuthSession may continue to reuse the established authentication fact, belong to an independent session contract. A successful authentication does not imply that every context must create a long-lived AuthSession.

SoulAuth can therefore represent two different facts at once: an Actor did in fact authenticate at some point in the past, and the current request may or may not still be allowed to rely on that authentication. The first is an already established authentication fact. The second remains constrained by freshness, session state, and the current trust context. \textbf{``This Actor proved who it was before'' does not imply ``every context should continue to trust that proof now.''}

\subsection{Identity Continuity and Authentication Trust Continuity}\label{identity-continuity-and-authentication-trust-continuity}

Once the lifecycles of Actor, Credential, Client, IdentityBinding, and AuthSession have been separated, the runtime must answer a further question: \textbf{when one object changes, what should happen to authentication facts established in the past?}

Two kinds of continuity must be distinguished. \textbf{Identity continuity} asks whether the system is still dealing with the same Actor after a legitimate change to Credential, session, Client, or other peripheral state. \textbf{Authentication trust continuity} asks whether an earlier Authentication Result or AuthSession may still be relied upon after such a change.

The answer cannot be derived by simple object equivalence. Revoking a Credential does not make ActorIdentity disappear, but whether an authentication state previously established using that Credential remains usable must be determined by explicit lifecycle-propagation rules. Suspending or retiring an Actor likewise does not require every Credential, AuthSession, or historical audit record to disappear from storage, but those state changes affect which authentication facts may continue to be accepted. Logout similarly terminates a particular AuthSession or scope of authentication continuity; it is not automatically equivalent to Credential Revocation or Actor Retirement.

This lets SoulAuth preserve identity stability and trust revocability at the same time. An Actor may remain the same Actor even when a previous authentication no longer satisfies current conditions. Conversely, revoking an AuthSession does not mean the system has lost the ActorIdentity or its historical authentication facts.

This distinction is especially important for long-running AIActors. Credential Rotation, runtime updates, and AuthSession re-establishment may all be normal parts of their lifecycle. If every change in authentication state created a new Actor, identity continuity would collapse. If every successful authentication remained valid forever, the system would lose control over current trust.

At the runtime level, the relationship can be summarized as follows: \textbf{ActorIdentity can persist across authentication and AuthSession lifecycles, while each authentication is valid only within its own time, context, and trust conditions.}

\begin{longtable}[]{@{}
  >{\raggedright\arraybackslash}p{(\columnwidth - 6\tabcolsep) * \real{0.2500}}
  >{\raggedright\arraybackslash}p{(\columnwidth - 6\tabcolsep) * \real{0.2500}}
  >{\raggedright\arraybackslash}p{(\columnwidth - 6\tabcolsep) * \real{0.2500}}
  >{\raggedright\arraybackslash}p{(\columnwidth - 6\tabcolsep) * \real{0.2500}}@{}}
\toprule\noalign{}
\begin{minipage}[b]{\linewidth}\raggedright
Change
\end{minipage} & \begin{minipage}[b]{\linewidth}\raggedright
Does ActorIdentity Change?
\end{minipage} & \begin{minipage}[b]{\linewidth}\raggedright
Existing Authentication Trust
\end{minipage} & \begin{minipage}[b]{\linewidth}\raggedright
Historical Attribution
\end{minipage} \\
\midrule\noalign{}
\endhead
\bottomrule\noalign{}
\endlastfoot
Credential rotation & No & Re-evaluate under the authentication contract and current context & Unchanged \\
Credential revocation & No & Related trust may become invalid & Unchanged \\
Client change & No & Evaluate under protocol and authentication context & Unchanged \\
AuthSession logout / expiry & No & The corresponding session ends & Unchanged \\
Runtime update / migration & No & Evaluate under the applicable trust contract & Unchanged \\
IdentityBinding revocation & No & The relevant cross-domain relationship ceases to apply & Existing history is not rewritten \\
Actor suspension & No & No new authentication trust beyond what current state permits & Unchanged \\
Actor retirement & ActorIdentity is not reused & No new valid authentication & Historical attribution is retained \\
\end{longtable}

\textbf{Identity continuity answers ``is this still the same Actor?'' Authentication trust continuity answers ``may a previously established authentication fact still be relied upon now?'' The former does not imply the latter.}

\begin{figure}[htbp]
\centering
\includegraphics{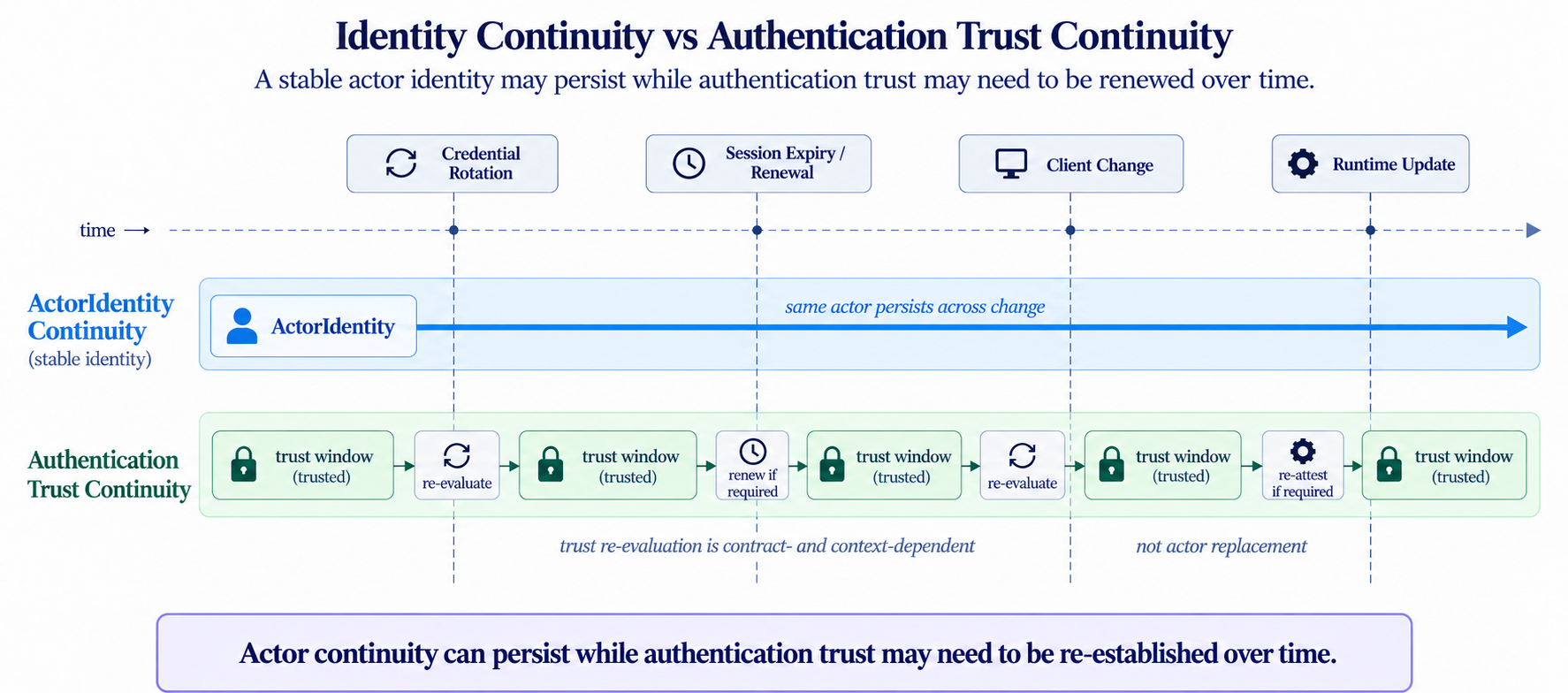}
\caption{Figure 6 \textbar{} Identity Continuity vs Authentication Trust Continuity}
\end{figure}

\section{Rust Reference Implementation}\label{rust-reference-implementation}

\subsection{From Actor-native Identity Architecture to a Real System}\label{from-actor-native-identity-architecture-to-a-real-system}

A conceptual model alone cannot establish that Actor-native Identity can serve as real identity infrastructure. The SoulAuth reference implementation therefore has a specific purpose: to preserve, in real software, the object boundaries among \texttt{ActorIdentity}, HumanAccount, Credential, Client, IdentityBinding, and AuthSession, together with the responsibility boundaries among identity, authentication, protocol, audit, and persistence.

SoulAuth implements this architecture in Rust. Rust is an engineering vehicle here, not a theoretical precondition for Actor-native Identity. The architecture first defines object semantics, object boundaries, and responsibility boundaries; the implementation then carries those constraints into data structures, runtime logic, protocol entry points, and persistence relations. The role of the reference implementation is to turn an established identity model into compilable, deployable, inspectable software, not to let implementation details redefine the architecture.

This chapter therefore focuses on \textbf{how the key structural relations of Actor-native Identity are preserved in a real runtime}, rather than cataloging the technical components used by SoulAuth.

\subsubsection{Implementation Overview}\label{implementation-overview}

\begin{longtable}[]{@{}
  >{\raggedright\arraybackslash}p{(\columnwidth - 2\tabcolsep) * \real{0.5000}}
  >{\raggedright\arraybackslash}p{(\columnwidth - 2\tabcolsep) * \real{0.5000}}@{}}
\toprule\noalign{}
\begin{minipage}[b]{\linewidth}\raggedright
Item
\end{minipage} & \begin{minipage}[b]{\linewidth}\raggedright
Description
\end{minipage} \\
\midrule\noalign{}
\endhead
\bottomrule\noalign{}
\endlastfoot
\textbf{Reference Implementation} & SoulAuth \texttt{v0.1.0} \\
\textbf{Implementation Language} & Rust \\
\textbf{Open-source Repository} & \url{https://github.com/TrantorLabs/SoulAuth} \\
\textbf{License} & Apache-2.0 \\
\textbf{Evaluation Boundary} & Fixed source revision; see Artifact Availability \\
\end{longtable}

The public reference implementation includes the implementation surfaces most directly related to the paper's argument: ActorIdentity, HumanAccount, native AIActor authentication, Client, AuthSession, OIDC, audit, and Architecture Conformance. The purpose of this overview is not to enumerate release metadata or test counts, but to establish that Actor-native Identity has been realized in a real, runnable, and externally inspectable software system. The exact revision, archive checksum, and CI entry point are consolidated in the Artifact Availability section.

\subsection{From Identity Model to Implementation Objects}\label{from-identity-model-to-implementation-objects}

At the implementation layer, Actor-native Identity first appears as real separation among identity objects. ActorIdentity remains the canonical continuity boundary, while HumanAccount, Credential, Client, IdentityBinding, and AuthSession each retain their own state and lifecycle rather than being recombined into a single User object.

The architecture-to-implementation mapping can be summarized as follows:

\begin{longtable}[]{@{}
  >{\raggedright\arraybackslash}p{(\columnwidth - 2\tabcolsep) * \real{0.5000}}
  >{\raggedright\arraybackslash}p{(\columnwidth - 2\tabcolsep) * \real{0.5000}}@{}}
\toprule\noalign{}
\begin{minipage}[b]{\linewidth}\raggedright
Architectural Semantics
\end{minipage} & \begin{minipage}[b]{\linewidth}\raggedright
Relation the Implementation Must Preserve
\end{minipage} \\
\midrule\noalign{}
\endhead
\bottomrule\noalign{}
\endlastfoot
\texttt{ActorIdentity} is the canonical continuity boundary & Authentication subject, Credential ownership, AuthSession subject, and attribution ultimately resolve to a specific ActorIdentity \\
\texttt{HumanAccount\ \ensuremath{\neq}\ ActorIdentity} & Human-specific account state remains separate from the canonical continuity boundary \\
\texttt{Credential\ \ensuremath{\neq}\ ActorIdentity} & Credential has independent state and lifecycle and an explicit relationship to ActorIdentity \\
\texttt{Client\ \ensuremath{\neq}\ Actor} & Protocol-level Client identity and authenticated subject are resolved separately \\
\texttt{IdentityBinding\ \ensuremath{\neq}\ ActorIdentity} & Cross-domain relationship exists independently and does not overwrite either identity object \\
\texttt{AuthSession\ \ensuremath{\neq}\ ActorIdentity} & Session state may reference ActorIdentity but does not own the Actor lifecycle \\
\end{longtable}

This mapping does not require the source tree to mirror the architecture diagram one-to-one. One logical responsibility may be implemented by several Rust modules, and one module may contain several internal functions. What must remain stable is the direction of semantic dependence: the canonical continuity boundary must not collapse back into HumanAccount, Credential, Client, or AuthSession, and protocol objects must not acquire subject-defining authority for reasons of implementation convenience.

The Human and AIActor implementation paths follow the same rule. Humans may retain account state and Human-specific authentication mechanisms. AIActors may have identity and authentication relationships independent of HumanAccount. The two need not share Credentials or protocol flows, but both must ultimately resolve to ActorIdentity as the common identity layer.

\begin{figure}[htbp]
\centering
\includegraphics{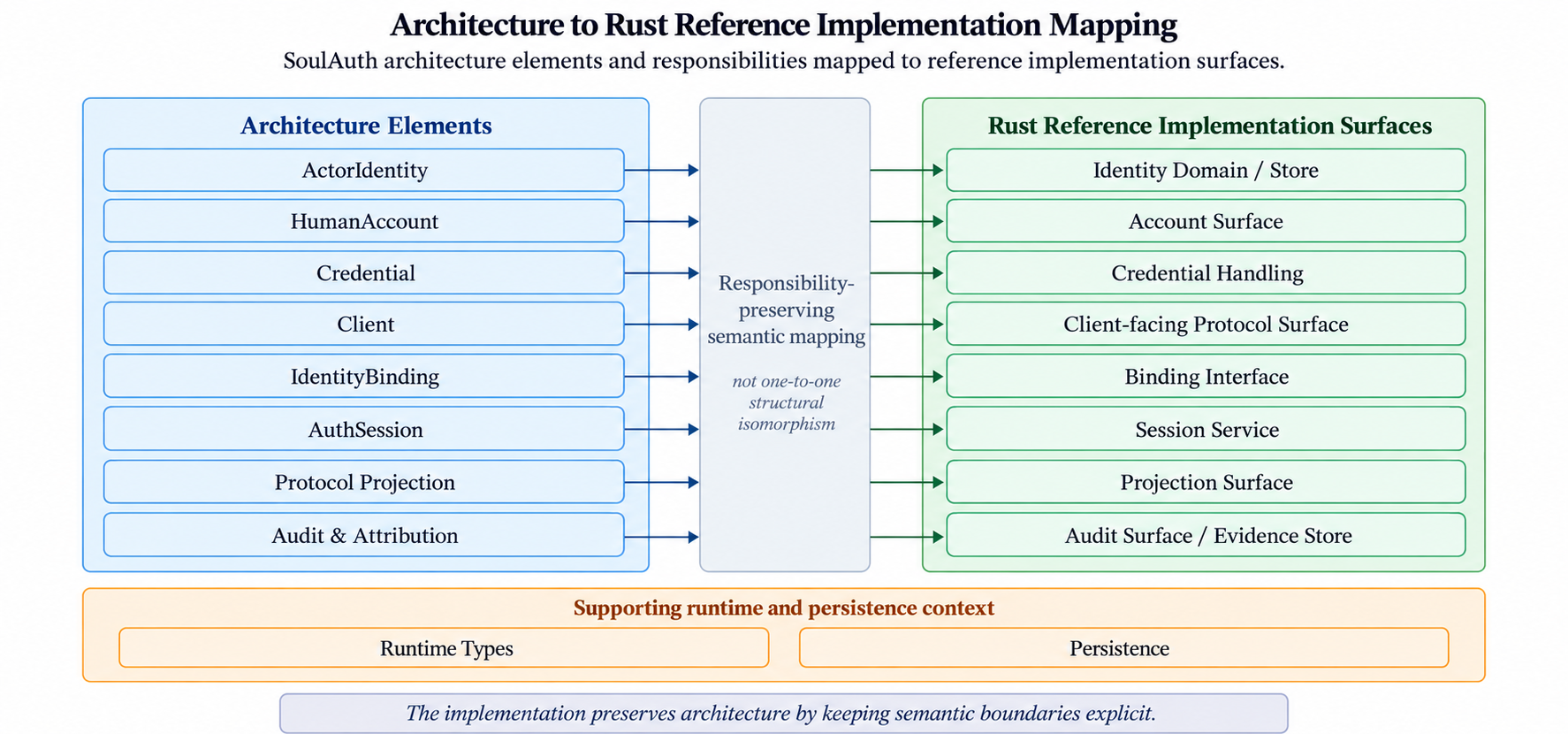}
\caption{Figure 7 \textbar{} Architecture to Rust Reference Implementation Mapping}
\end{figure}

\subsection{Runtime Realization of Key Architectural Boundaries}\label{runtime-realization-of-key-architectural-boundaries}

Evaluating whether the reference implementation preserves Actor-native Identity does not require enumerating every feature. It requires examining the runtime boundaries that most directly determine whether the intended semantics still hold.

The first boundary is the separation between \textbf{ActorIdentity and Credential}. Credentials may be created, rotated, revoked, or invalidated, but those changes must not redefine ActorIdentity. The authentication runtime consumes a Credential or another Authentication Source and establishes an Authentication Result about a particular ActorIdentity. Credential remains responsible for ``how the subject proves itself''; ActorIdentity remains responsible for ``who the subject is.''

The second boundary is the separation between \textbf{Client and Actor}. An OAuth / OIDC Client has its own \texttt{client\_id}, registration data, and protocol state, but Client authentication must not automatically become Actor authentication. Protocol entry points must separately process ``which Client is participating in the protocol?'' and ``which ActorIdentity is currently being authenticated?''

The third boundary is \textbf{native AIActor authentication}. Support for AIActors cannot be only a new type label. An AIActor must be able to enter an authentication path under its own ActorIdentity without depending on HumanAccount. A long-lived authentication capability and the proof submitted in one specific request operate on different timescales. Successful verification produces an Authentication Result; it does not turn the key, proof, or AuthSession itself into the Actor.

The fourth boundary is the separation between \textbf{protocol projection and ActorIdentity}. OIDC \texttt{sub}, Token, Claims, UserInfo, and other relying-party representations may carry established identity and authentication facts, but they remain protocol projections. Protocol implementation must preserve stable mappings without allowing a Token, Claim, Client identifier, or mutable Profile attribute to become the Source of Truth for ActorIdentity.

Persistence follows the same rule. The database stores domain state, but a concrete Schema does not own the definition of the upper-level ontology. \texttt{Persistence\ Schema\ \ensuremath{\neq}\ Canonical\ Ontology} is therefore not only an architectural statement; it is an engineering boundary that must survive implementation evolution.

Together, these checks ask a single question: \textbf{do the identity objects and responsibilities separated in the architecture remain semantically separated at runtime?}

\subsection{Rust, Open Source, and Artifact Inspectability}\label{rust-open-source-and-artifact-inspectability}

Rust contributes to SoulAuth primarily at the implementation layer. Identity and authentication infrastructure must continuously handle Credentials, Tokens, cryptographic material, untrusted protocol input, and concurrent state. Memory safety, ownership constraints, and an explicit type system can reduce some classes of low-level implementation risk and provide stronger engineering expression for object and state boundaries.

These language-level properties do not replace the upper-level architecture. Rust cannot determine whether an IdentityBinding is semantically correct, nor can it automatically ensure that authentication has not crossed the Authority boundary. Rust therefore strengthens the implementation; it does not prove the architecture correct.

Open-source code gives architectural claims an engineering artifact that external reviewers can inspect. Source code, Schemas, protocol implementation, machine-readable contracts, and tests can be examined independently to determine whether boundaries involving ActorIdentity, Credential, Client, AIActor authentication, AuthSession, and attribution actually exist in the software.

For v0.1.0, this paper reports only implementation and evaluation facts that are directly relevant to the research claims and bound to a fixed source revision; these results are not generalized to later states of \texttt{main}. The exact commit, versioned release archive, and checksum are consolidated in the Artifact Availability section, and later changes to \texttt{main} fall outside this paper's evidence boundary.

Rust and open source therefore play different roles in this paper: \textbf{Rust carries Actor-native Identity into a real implementation, while open source makes the architecture-to-implementation mapping independently inspectable.}

\section{Architecture Conformance and Artifact Evidence}\label{architecture-conformance-and-artifact-evidence}

\subsection{From Architectural Claims to Implementation Evidence}\label{from-architectural-claims-to-implementation-evidence}

For SoulAuth, functional success alone is not sufficient evidence that the Actor-native Identity architecture has been realized. A system may continue to authenticate, create AuthSessions, or issue Tokens while internally collapsing ActorIdentity back into User, treating Client as Actor, or allowing an Authentication Result to acquire Authority that does not belong to the identity layer. Such changes may not appear as ordinary functional failures, yet they constitute architectural drift.

SoulAuth therefore does not treat ``the system runs'' as sufficient evidence that the architecture has been realized. Instead, the paper maps key architectural claims to concrete implementation surfaces and verification evidence:

\begin{quote}
\textbf{Architectural Claim → Implementation Surface → Verification Evidence}
\end{quote}

This chain is also the testable interface of Philosophical Engineering in SoulAuth: \textbf{once a normative judgment about subjecthood has been translated into explicit invariants, the implementation can be observed to satisfy, partially satisfy, or violate those constraints at concrete engineering surfaces.} Conformance testing therefore asks not only whether a feature works, but whether the implementation realizes the subject structure the architecture claims to adopt.

Different claims require different inspection surfaces. Object separation may be reflected in types, Schemas, and repositories. Protocol boundaries may be reflected in runtime behavior and machine-readable contracts. Lifecycle constraints may be tested through state transitions and negative tests. No single test can prove every architectural claim. What matters is that \textbf{evidence must be claim-specific; ``the tests pass'' cannot substitute for verification of architectural semantics.}

Machine-readable contracts form an intermediate layer. OpenAPI, registries, and similar contracts can describe a public surface precisely, but they do not thereby own the upstream ontology. Human-readable semantics, machine-readable contracts, runtime behavior, and tests should agree where their scopes overlap. If they make conflicting claims about the same architectural property, that conflict should be treated as a conformance defect.

\subsection{Conformance Verification of Key Actor-native Identity Invariants}\label{conformance-verification-of-key-actor-native-identity-invariants}

The verification targets for Architecture Conformance come directly from the core invariants established by the Actor-native Identity model; testing does not invent a second rule set. The paper focuses on boundaries whose failure would cause SoulAuth to collapse back into Account-centric, Credential-centric, or Client-centric identity.

\begin{longtable}[]{@{}
  >{\raggedright\arraybackslash}p{(\columnwidth - 4\tabcolsep) * \real{0.3333}}
  >{\raggedright\arraybackslash}p{(\columnwidth - 4\tabcolsep) * \real{0.3333}}
  >{\raggedright\arraybackslash}p{(\columnwidth - 4\tabcolsep) * \real{0.3333}}@{}}
\toprule\noalign{}
\begin{minipage}[b]{\linewidth}\raggedright
Architectural Claim
\end{minipage} & \begin{minipage}[b]{\linewidth}\raggedright
Primary Implementation Surface
\end{minipage} & \begin{minipage}[b]{\linewidth}\raggedright
Typical Verification Evidence
\end{minipage} \\
\midrule\noalign{}
\endhead
\bottomrule\noalign{}
\endlastfoot
\textbf{AIActor does not depend on HumanAccount} & Creation and association logic for ActorIdentity and HumanAccount & Schema, runtime behavior, conformance tests \\
\textbf{ActorIdentity \ensuremath{\neq} Credential} & Credential ownership and lifecycle & Type / Schema checks, lifecycle tests \\
\textbf{Client \ensuremath{\neq} Actor} & Client registration and Subject resolution & Protocol runtime, negative tests \\
\textbf{Credential Rotation \ensuremath{\neq} Actor Replacement} & Credential rotation path & State-transition tests \\
\textbf{Authentication \ensuremath{\neq} Authority} & Authentication Result and Authority boundary & Runtime checks, negative conformance checks \\
\textbf{Historical attribution binds to stable ActorIdentity} & Audit and attribution path & Audit tests, integrity evidence \\
\textbf{Protocol Subject does not define ActorIdentity in reverse} & OIDC / Claims / Subject mapping & Contract, runtime, conformance tests \\
\end{longtable}

These checks can detect semantic regressions that ordinary functional tests may miss. AIActor authentication may still succeed, for example, while the runtime still requires a HumanAccount before the AIActor can become a first-class identity subject. In that case, the Actor-native Identity model has already failed. Likewise, OAuth / OIDC may complete normally while an implementation interprets \texttt{client\_id} directly as the Subject of the authenticated Actor, violating \texttt{Client\ \ensuremath{\neq}\ Actor}.

Credential rotation provides another example. Successfully switching to a new Credential proves only that rotation functionality exists. Architecture Conformance must additionally verify that ActorIdentity remains unchanged through the transition. \texttt{Authentication\ \ensuremath{\neq}\ Authority} likewise requires negative evidence: producing an Authentication Result must not simultaneously grant downstream application, governance, or execution Authority.

Architecture Conformance therefore does not ask merely whether the system completed an operation. It asks:

\begin{quote}
\textbf{Did the system complete the operation using the correct identity relationships?}
\end{quote}

For systems in which Humans and long-lived AIActors coexist, this distinction has direct implications for responsibility. Functionally successful authentication with incorrect subject attribution is not merely internal semantic drift; it undermines the identity foundation on which downstream accountability, audit, and governance rely.

\subsection{Conformance Results on the Fixed v0.1.0 Revision}\label{conformance-results-on-the-fixed-v0.1.0-revision}

To test whether Actor-native Identity has actually entered the implementation, we fix the evaluation target to a specific SoulAuth \texttt{v0.1.0} source revision and use its public CI to evaluate the build, unit tests, integration tests, and Architecture Conformance. On that revision, \textbf{188 unit tests, 61 active conformance checks, and 355 integration assertions passed}. At the same time, the test suite continues to track \textbf{nine conformance checks that remain unsatisfied}. We therefore do not treat ``the system runs'' or ``all active checks pass'' as evidence of full architectural conformance; the current implementation is therefore reported as exhibiting \textbf{partial conformance}.

At the level of the paper's central claims, the evaluation can be summarized as follows:

\begin{longtable}[]{@{}
  >{\raggedright\arraybackslash}p{(\columnwidth - 2\tabcolsep) * \real{0.5000}}
  >{\raggedright\arraybackslash}p{(\columnwidth - 2\tabcolsep) * \real{0.5000}}@{}}
\toprule\noalign{}
\begin{minipage}[b]{\linewidth}\raggedright
Core Architectural Property
\end{minipage} & \begin{minipage}[b]{\linewidth}\raggedright
Observation in v0.1.0
\end{minipage} \\
\midrule\noalign{}
\endhead
\bottomrule\noalign{}
\endlastfoot
\textbf{Humans and AIActors can both be first-class identity subjects} & Realized \\
\textbf{Client, AuthSession, and protocol projections do not replace ActorIdentity} & Realized \\
\textbf{IdentityBinding represents a relation, not identity equivalence} & Realized \\
\textbf{Authentication does not automatically create downstream Authority} & Realized \\
\textbf{Full lifecycle separation between Credential and ActorIdentity} & Partially realized \\
\textbf{Historical attribution is uniformly anchored to ActorIdentity} & Partially realized \\
\end{longtable}

These results are not intended as an implementation scorecard; their significance is that they directly test whether the normative claims developed earlier have been realized in the software.

First, an AIActor can be created and authenticate without depending on a HumanAccount. This shows that equal first-class status as identity subjects for Humans and long-lived AIActors is not confined to the object model; it is present in the actual authentication path.

Second, separation between Credential and ActorIdentity is only partially realized. AIActor key lifecycles can already change independently without replacing the Actor, but Human-side Credentials have not yet converged on a fully unified independent model. The invariant \texttt{ActorIdentity\ \ensuremath{\neq}\ Credential} therefore has concrete implementation support, but it is not yet closed across every path.

Third, historical attribution is also only partially realized. The audit subsystem already provides independent security and integrity mechanisms, but some historical events still rely on Human-oriented \texttt{user\_id} references rather than uniformly resolving to stable ActorIdentity. This result illustrates why Architecture Conformance matters: \textbf{a system can pass functional tests and still fail to realize an adopted subject structure completely at a particular engineering surface.}

The remaining conformance gaps continue to be tracked in the public artifact. The paper keeps only the observations that directly bear on its research claims; complete test identifiers, execution logs, and engineering diagnostics are available in the fixed public artifact.

\subsection{Inspectable Artifacts and Evidence Boundary}\label{inspectable-artifacts-and-evidence-boundary}

Architecture evidence is inspectable only when it is bound to a specific software artifact. This paper fixes the evaluation boundary to a specific SoulAuth \texttt{v0.1.0} source revision so that paper claims, implementation surfaces, test targets, and observed results all refer to the same artifact. The exact commit, release archive, checksum, and public CI entry point are consolidated in the Artifact Availability section.

The scope of the evidence must match the scope of the claim. A lifecycle test may support the statement that Credential Rotation did not replace ActorIdentity, but it cannot prove that the entire identity system is secure. Architecture Conformance likewise does not automatically imply third-party protocol certification. Each evidence item carries only the burden of proof appropriate to the claim it supports.

SoulAuth's artifact evidence can therefore be summarized as:

\begin{quote}
\textbf{Architectural Claim → Implementation Surface → Verification Evidence → Public Artifact}
\end{quote}

\begin{figure}[htbp]
\centering
\includegraphics{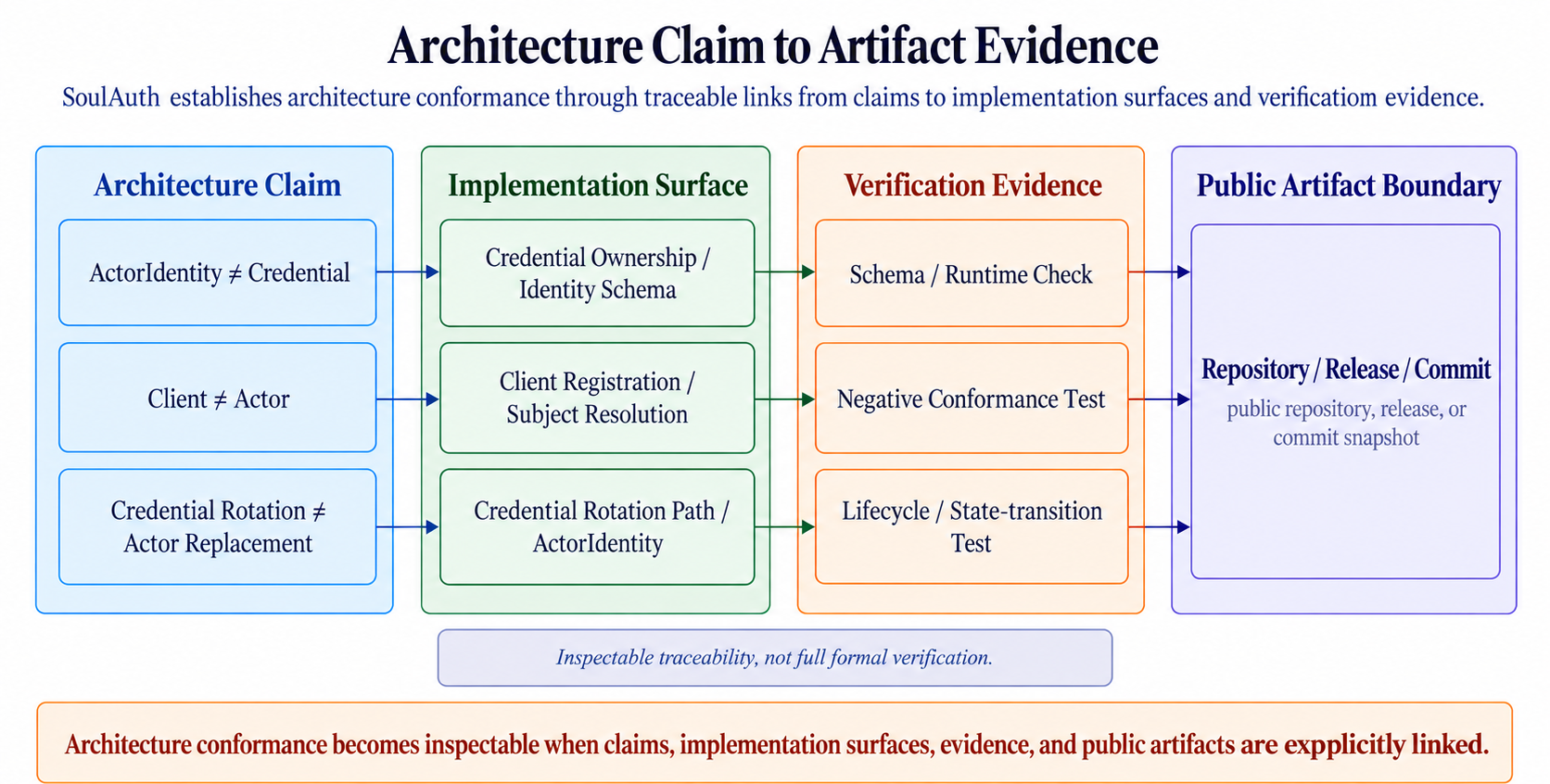}
\caption{Figure 8 \textbar{} Architecture Claim to Artifact Evidence}
\end{figure}

This chain does not make the generalized claim that ``tests pass, therefore the system is correct.'' Its purpose is narrower and more inspectable: to determine whether the core Actor-native architectural claims in this paper have traceable engineering counterparts in the public reference implementation.

For an architecture-and-reference-implementation paper, this kind of inspectability is more important than importing an entire release-governance system into the main text. It creates an explicit connection among the paper's identity model, system architecture, and implementation artifact while allowing limitations, partial implementation state, and future evolution to be stated directly.

\section{Security Analysis}\label{security-analysis}

\subsection{Threat Model and Identity Misattribution}\label{threat-model-and-identity-misattribution}

SoulAuth's security goal is not merely to prevent failed authentication. As an identity and authentication security architecture, SoulAuth focuses on four security properties directly related to Actor-native Identity: \textbf{ActorIdentity integrity, authentication integrity, identity-relationship integrity, and historical-attribution integrity}. Correct subject attribution is a common requirement across all four. A proof can be valid both cryptographically and at the protocol level and still be attributed to the wrong persistent subject. Protocol, administrative, or infrastructure attacks enter the analysis in this chapter only when they can violate one of these properties. For Actor-native identity infrastructure, a more serious failure mode is therefore one in which \textbf{authentication succeeds, but the proof is attributed to the wrong ActorIdentity.}

Identity misattribution is thus the central security concern of this chapter. Failing to resolve an Actor is not the same as identifying Actor A as Actor B. The first typically means that an identity relationship has not been established. The second means the system has already established incorrect trust and may go on to create an AuthSession, generate protocol projections, record audit attribution, or deliver the wrong identity fact to downstream systems.

The threat model does not classify Humans as trusted and AIActors as untrusted, or vice versa. ActorKind is not a trust level, and AIActor is not a threat category. Threats are analyzed according to the capabilities available to an attacker or fault source, the trust boundaries those capabilities can cross, and whether they are sufficient to corrupt identity or authentication relationships. Complete compromise of the system's root of trust is treated as foundational trust failure rather than an ordinary authentication failure.

To make this boundary reviewable, the threat model states the following attacker capabilities, trust assumptions, and non-goals explicitly:

\begin{longtable}[]{@{}
  >{\raggedright\arraybackslash}p{(\columnwidth - 4\tabcolsep) * \real{0.3333}}
  >{\raggedright\arraybackslash}p{(\columnwidth - 4\tabcolsep) * \real{0.3333}}
  >{\raggedright\arraybackslash}p{(\columnwidth - 4\tabcolsep) * \real{0.3333}}@{}}
\toprule\noalign{}
\begin{minipage}[b]{\linewidth}\raggedright
Attack / Fault Capability
\end{minipage} & \begin{minipage}[b]{\linewidth}\raggedright
Scope
\end{minipage} & \begin{minipage}[b]{\linewidth}\raggedright
Security Consequence of Interest
\end{minipage} \\
\midrule\noalign{}
\endhead
\bottomrule\noalign{}
\endlastfoot
Possession or theft of a valid Credential & In scope & Whether authentication is misattributed to another ActorIdentity \\
Control of a registered Client & In scope & Whether Client is incorrectly elevated into Actor \\
Submission of stale, replayed, or context-mismatched authentication evidence & In scope & Authentication integrity, freshness, and authentication trust continuity \\
Binding attacker-controlled verification material to another Actor & In scope & Credential binding substitution \\
Exploiting similar external Subjects / emails to induce cross-domain identity merging & In scope & IdentityBinding and identity-relationship integrity \\
Exploiting revocation delay, stale cache, or restored state & In scope & Whether invalidated trust is incorrectly revived as current trust \\
Downstream interpretation of authentication success as unlimited Authority & Analyzed at the integration boundary & Failure of the Authentication / Authority boundary \\
Complete compromise of root signing keys, trusted persistent state, and the core runtime & Foundational trust failure / not an ordinary authentication threat & Not modeled as a general identity-relation failure \\
Complete compromise of the operating system, compiler, or cryptographic primitives & Outside the trusted computing base assumptions & Out of scope \\
Whether an AIActor is conscious or has legal personhood & Non-goal & Outside the threat model \\
\end{longtable}

The paper therefore assumes that the core runtime performing identity verification, the cryptographic primitives it depends on, trusted key material, and canonical persistent state are not fully controlled by an attacker. Under those foundational assumptions, the security analysis asks whether the relationships among ActorIdentity, Credential, Client, IdentityBinding, AuthSession, protocol projection, and historical attribution remain correct.

An important security property of Actor-native Identity is that a serious failure need not stop the system from functioning. A system may still verify a valid signature, parse a structurally valid Binding, or issue a Token. If the system has established the wrong Actor, Credential ownership, or identity relationship, the relevant security property has already failed.

SoulAuth therefore must first ask not ``did authentication succeed?'' but:

\begin{quote}
\textbf{Did this authentication actually belong to the correct ActorIdentity?}
\end{quote}

To connect this risk directly to the invariants, architectural mechanisms, and engineering evidence introduced earlier, the following table lists five representative identity failure modes:

\begin{longtable}[]{@{}
  >{\raggedright\arraybackslash}p{(\columnwidth - 8\tabcolsep) * \real{0.2000}}
  >{\raggedright\arraybackslash}p{(\columnwidth - 8\tabcolsep) * \real{0.2000}}
  >{\raggedright\arraybackslash}p{(\columnwidth - 8\tabcolsep) * \real{0.2000}}
  >{\raggedright\arraybackslash}p{(\columnwidth - 8\tabcolsep) * \real{0.2000}}
  >{\raggedright\arraybackslash}p{(\columnwidth - 8\tabcolsep) * \real{0.2000}}@{}}
\toprule\noalign{}
\begin{minipage}[b]{\linewidth}\raggedright
Failure Mode
\end{minipage} & \begin{minipage}[b]{\linewidth}\raggedright
Incorrect Relation
\end{minipage} & \begin{minipage}[b]{\linewidth}\raggedright
Violated Invariant / Boundary
\end{minipage} & \begin{minipage}[b]{\linewidth}\raggedright
SoulAuth Prevention Mechanism
\end{minipage} & \begin{minipage}[b]{\linewidth}\raggedright
Implementation / Evidence Surface
\end{minipage} \\
\midrule\noalign{}
\endhead
\bottomrule\noalign{}
\endlastfoot
Credential becomes the subject & Credential = Actor & I2 / I5 & Credential has an independent lifecycle and explicit ownership by ActorIdentity & Credential ownership, Schema, lifecycle / conformance evidence \\
Client becomes the subject & \texttt{client\_id} = Actor & I3 / I6 & Client and authenticated Actor are resolved separately & Client registration, Subject resolution, negative conformance evidence \\
Runtime instance becomes the subject & Runtime Instance = Actor & Canonical continuity boundary & Runtime changes do not redefine ActorIdentity & Identity Domain, lifecycle / state-transition evidence \\
Binding causes identity merging & Similar external identifier = same Actor & I4 / Relationship \ensuremath{\neq} Identity Equivalence & Explicit IdentityBinding and trust contract & IdentityBinding interface, negative / integrity evidence \\
Authentication success creates Authority & Authentication Success = Authority & I8 & Identity, Authentication, and Authority remain separate & Authentication Result boundary, negative conformance evidence \\
\end{longtable}

Section 7.3 summarizes the implementation-level observations most directly relevant to these failure modes; detailed test-level evidence remains in the fixed public artifact.

\subsection{Identity Relationships and Authentication Integrity}\label{identity-relationships-and-authentication-integrity}

Once ActorIdentity, Credential, Client, IdentityBinding, and protocol artifacts are separated, security depends not only on whether each object is individually valid, but also on whether the relationships among them are correct.

Credential security means more than keeping secret material confidential. It also means ensuring that the Credential or verification material remains bound to the correct ActorIdentity. Even without stealing Actor A's original Credential, an attacker who can bind attacker-controlled verification material to Actor A may produce authentication that passes cryptographic verification but is attributed to the wrong subject. Credential theft and Credential binding substitution are therefore distinct threats.

Cryptographic verification itself is likewise only a necessary condition. A mathematically valid signature does not automatically establish that the proof belongs to the correct Actor, the correct Challenge, or the correct temporal context. Authentication evidence must also satisfy Actor binding, context, freshness, and related constraints.

The same class of relationship-integrity problem appears at the Client and federation boundaries. Successful Client authentication proves only that a Client satisfies its protocol contract; it does not automatically establish Actor authentication. Likewise, an external Subject must be interpreted together with an explicit identity source, Issuer, and IdentityBinding before it can resolve to a local ActorIdentity. A valid Client, external authentication result, or protocol artifact can still cause identity misattribution if it enters the wrong identity relationship or protocol context.

Actor-native Identity security must therefore verify not only the object itself, but also:

\begin{quote}
\textbf{who owns it, whom it represents, and why that identity relationship is valid in the current context.}
\end{quote}

\subsection{Authority and Temporal Trust Boundaries}\label{authority-and-temporal-trust-boundaries}

Even when SoulAuth authenticates the correct Actor and all relevant identity relationships are valid, an authentication fact may only be used within its proper scope.

The first is the \textbf{Authority boundary}. An Authentication Result states that an ActorIdentity has been proven under particular conditions. It does not automatically create application, governance, or execution Authority. If a relying system interprets authentication success as final authorization for business or governance actions, the Authority boundary has failed. Administrative Authority is likewise scoped: the ability to manage one class of security state does not imply unrestricted Authority over every identity, key, or permission.

The second boundary is \textbf{time}. ActorIdentity may persist while Credential, IdentityBinding, Authentication Result, and AuthSession change over time. Current state must not retroactively redefine authentication and attribution facts that already occurred. Conversely, trust that was once valid but later invalidated must not be revived as current trust merely because of rollback, replica lag, or recovery.

Therefore:

\begin{quote}
\textbf{Current identity state may change, but historical attribution must not be rewritten by current state; trust that was valid in the past is not necessarily valid now.}
\end{quote}

SoulAuth's security analysis can thus be summarized in three layers. The system must authenticate the correct Actor. Credentials, Clients, Bindings, and protocol artifacts must have the correct relationship to that Actor. And even when the identity and authentication facts are themselves correct, they may only be used within the correct Authority scope and time window.

In other words, SoulAuth must know not only whether an object is valid, but also \textbf{who owns it, whom it represents, and within what scope and time the resulting fact remains valid.}

\section{Related Work and Discussion}\label{related-work-and-discussion}

SoulAuth sits at the intersection of digital identity, Workload Identity, AI agent identity, continuity architectures, and responsibility attribution. Existing work has already developed mature or rapidly evolving approaches around Principal, Protocol Subject, Workload Identity, DID, Agent authentication, delegation, and accountability. SoulAuth builds on these foundations and narrows the research question to a structural one: \textbf{when Humans and long-lived AIActors share the same identity infrastructure, which object should carry the canonical continuity boundary?}

\subsection{Principal, Protocol Identity, and Workload Identity}\label{principal-protocol-identity-and-workload-identity}

Computer security already has general abstractions for non-Human subjects. Lampson, Abadi, Burrows, and Wobber used \texttt{Principal} as a general subject concept in their theory of authentication for distributed systems and analyzed names, communication channels, roles, delegation, access control, and revocation \cite{ref1}. SoulAuth therefore builds on the general Principal abstraction rather than redefining Principal itself.

OAuth and OpenID Connect have established mature semantics for protocol participants, Authentication Results, and Subjects. OpenID Connect explicitly distinguishes Client from End-User Subject and specifies \texttt{iss\ +\ sub} as a stable subject-identifier combination on which a Relying Party may depend, whereas Claims such as email and username do not receive the same stability guarantee \cite{ref2}. SoulAuth retains these standard protocol roles rather than changing their semantics. Its concern lies one level upstream: \textbf{which persistent identity object in the identity domain should a protocol-level Subject be a projection of?} In SoulAuth, OIDC \texttt{sub}, Claims, Token, and UserInfo are protocol projections of identity facts rather than ActorIdentity itself.

Identity infrastructure for machine subjects is likewise mature. SPIFFE defines SPIFFE ID, SVID, Workload API, and Trust Domain, providing verifiable identity and federation for Workloads running across heterogeneous environments \cite{ref3}. IETF WIMSE extends this line of work toward Workload identifiers, Credentials, and authentication mechanisms in multi-system environments, with Workload as the identity subject used across multiple runtime instances \cite{ref4}. SoulAuth therefore does not need to re-solve the question of whether a non-Human computational entity can have a verifiable identity. It instead asks which object should carry continuity for a long-lived Actor.

This line of work is now extending directly toward AI Agents. Individual Internet-Drafts in the WIMSE space have begun discussing independent AI Agent identities and Credentials. \texttt{WIMSE\ Applicability\ for\ AI\ Agents} proposes separating AI agent identity from device and user identity and discusses automated Credential management and the binding of AI agent identity to Owner Identity \cite{ref5}. Because the document is an individual Internet-Draft, we treat it as an emerging technical discussion rather than an IETF-endorsed standard.

The difference between SoulAuth and Workload Identity should therefore not be framed as a binary between ``machine identity'' and ``AI identity.'' It lies in the \textbf{canonical continuity boundary}. SPIFFE and WIMSE take Workload as their canonical identity subject, whereas SoulAuth takes ActorIdentity as its canonical identity subject. An AIActor may run inside an execution environment that has its own Workload Identity; the two identities are compatible. The former answers which Workload is currently running or invoking infrastructure. The latter answers whether authentication activity still belongs to the same persistent Actor after legitimate changes to runtime, Client, Credential, or AuthSession.

SoulAuth therefore keeps two identity semantics distinct:

\begin{quote}
\textbf{Workload Identity identifies the execution subject; ActorIdentity anchors the identity subject that persists across peripheral lifecycles.}
\end{quote}

\subsection{Persistent AI Agent Identity, Decentralized Identity, and Delegation}\label{persistent-ai-agent-identity-decentralized-identity-and-delegation}

Stable identity for long-lived Agents is already an explicit research direction. W3C Decentralized Identifiers (DIDs) allow a Subject to possess a decentralized identifier independent of any particular Verification Method and explicitly support Verification Method Rotation; changing a Key or Verification Method does not require changing the DID itself \cite{ref6}. \texttt{Credential\ Rotation\ \ensuremath{\neq}\ Identity\ Replacement} therefore has a direct precedent as a general identity principle, and SoulAuth incorporates that principle into the lifecycle invariants of Actor-native Identity.

Recent research applies DID / VC mechanisms directly to AI agent identity. Garzon et al.~assign long-lived DIDs to AI Agents and use Verifiable Credentials to establish cross-domain trust and authentication \cite{ref7}. AgentDID further combines DID, VC, and Challenge-Response to address self-managed AI agent identity, cross-system authentication, and runtime-state verification \cite{ref8}. These works already cover long-lived agent identity, public-key authentication, and cross-domain verifiable identity. SoulAuth therefore continues to move the research question upward, toward the boundary among identity objects themselves.

Another line of work extends continuity beyond Credential to runtime and topology. The Agent Identity URI Scheme uses \texttt{agent://} to decouple agent identity from network location so that an Agent can remain stably referenceable after migration across Provider, Instance, or organizational boundaries \cite{ref9}. Zhao and Zhao's Runtime-Independent Persistent Agents combine Identity, Memory, and a Versioned Software Body into a continuity-bearing substrate, treating Model, Harness, Host Server, and interaction surface as replaceable Deployment Bindings \cite{ref10}. These works are closely adjacent to SoulAuth because they likewise reject the equation of a runtime instance with a persistent Agent.

The research boundary remains different. Runtime-independent Agent Architecture asks \textbf{how the entire Agent persists}, with a continuity substrate spanning Identity, Memory, and Software Body. SoulAuth is deliberately limited to the identity layer and preserves:

\begin{quote}
\textbf{Identity Continuity \ensuremath{\neq} Mind or Behavioral Continuity.}
\end{quote}

SoulAuth can therefore record that events before and after runtime migration belong to the same ActorIdentity, but it does not decide whether Memory, Model, Policy, or Behavior still constitutes ``the same mind.''

A further major line of AI agent identity research concerns authorization and delegation. OpenID Foundation work on Agentic AI Identity treats agent identity, authentication, authorization, delegated authority, and AI Workload as distinct problems \cite{ref11}. South et al.~propose Authenticated Delegation, allowing a Human to delegate constrained authority to an Agent while preserving the relationship through Agent-specific Credentials, OAuth/OIDC compatibility, and an audit chain \cite{ref12}. AIP binds Identity, delegation scope, authorization, and Provenance into Invocation-Bound Capability Tokens and supports cross-transport verification for MCP, A2A, and HTTP \cite{ref13}. Agent Name Service (ANS) focuses primarily on secure Agent discovery, registration, renewal, and capability-aware resolution \cite{ref14}.

These works answer different questions: how an Agent proves identity, on whose behalf it acts, what capabilities it possesses, how other Agents discover it, and how invocation chains are recorded. SoulAuth keeps its own responsibility boundary prior to these mechanisms:

\begin{quote}
\textbf{Identity, Authentication, Authority, Delegation, and Discovery may be composed, but they should not be collapsed into one object.}
\end{quote}

A SoulAuth Authentication Result may serve as an input to authorization or delegation systems, but authentication does not create those powers by itself.

Against this background, SoulAuth focuses on object boundaries inside identity infrastructure: \textbf{ActorIdentity is the canonical continuity boundary, while HumanAccount, Credential, Client, IdentityBinding, AuthSession, protocol projection, and downstream Authority retain separate semantics and lifecycles.}

Existing work addresses different dimensions of this problem space. SoulAuth organizes these identity objects, lifecycles, and responsibility boundaries together as an \textbf{Actor-native Identity architecture}.

\subsection{Attribution, Applicability Boundaries, and Research Position}\label{attribution-applicability-boundaries-and-research-position}

As AI Agents begin performing tasks across systems, the relationship between identity and accountability is emerging as a distinct sociotechnical problem. Otsuka et al.~survey AI Identity as a persistent relation between the declared Agent and observed behavior and identify remaining gaps in persistence, agent identity integrity, recursive delegation accountability, and governance \cite{ref15}. Their notion of Identity is broader than SoulAuth's, spanning behavior, intent, and governance continuity. SoulAuth addresses only the more foundational identity and authentication sublayer.

Legal and institutional research further illustrates why ``who'' is a prerequisite for accountability. Arbel, Salib, and Goldstein argue in \emph{How to Count AIs} that before responsibility can be assigned, one must first answer ``Which AI Did It?'', and they distinguish thin identity, which links AI behavior to a Human Principal, from thick identity, which distinguishes persistent AI Entities from one another \cite{ref16}. He et al.~study a national agent-identity layer that relates agent identity to legal-subject binding, privacy-preserving re-identification, and ex post attribution \cite{ref17}. These works show that persistent identity is not merely an internal software-object problem; it can become infrastructure on which responsibility, audit, and institutional governance depend.

SoulAuth does not attempt to solve accountability directly. It provides a narrower and more foundational identity layer. In settings where Humans and long-lived AIActors share digital platforms, enterprise systems, or automated workflows, downstream accountability, delegation, and governance may still rest on the wrong subject if authentication cannot reliably answer ``which persistent subject was just proven,'' or if historical attribution changes when current Profile, Binding, Credential, or runtime state changes.

The sociotechnical significance of SoulAuth can therefore be summarized as follows:

\begin{quote}
\textbf{Accountability requires stable attribution, and stable attribution requires identity infrastructure to provide a ``who'' that does not drift when Credentials, Clients, sessions, or runtimes change legitimately.}
\end{quote}

This does not make identity-subject status equivalent to social or legal status. Humans and AIActors have equal first-class status as identity subjects in SoulAuth only in the sense that each can independently be identified, authenticated, and attributed within the identity domain; it does not imply equal Authority, capability, responsibility, or legal status.

ActorIdentity is not a globally universal identity either. It is first the canonical continuity boundary inside the SoulAuth identity domain. External systems may have their own Subjects, Principals, or Actor Ontologies. Relationships across identity domains must be established through explicit Bindings or contracts rather than inferred from identifier similarity.

Nor does every software process that uses AI need to become an AIActor. ActorIdentity is warranted only when a non-Human computational entity must maintain persistent identity across repeated authentication, Credentials, Clients, or runtime lifecycles and must become an independent subject of historical attribution.

The relationship between SoulAuth and existing work can therefore be summarized in three observations. First, SoulAuth builds on established work on general Principals, OAuth/OIDC, machine identity, and Workload Identity rather than replacing those systems. Second, it shares the broader problem of long-lived agent identity with work on DIDs, Persistent Agents, Agent delegation, and Agent discovery, while narrowing its scope to the identity and authentication layer. Third, its structural contribution is to choose a different \textbf{canonical continuity boundary}:

\begin{quote}
\textbf{Existing work can already provide identities for machines and AI Agents; SoulAuth asks a further question: when Humans and long-lived AIActors share one identity infrastructure, where should the canonical continuity boundary be placed?}
\end{quote}

SoulAuth answers:

\begin{quote}
\textbf{The canonical continuity boundary should be ActorIdentity, not Account, Credential, Client, Workload, AuthSession, Protocol Subject, or runtime instance.}
\end{quote}

This choice makes ActorIdentity a stable reference for authenticated subjects and historical attribution while allowing Account, Credential, Client, Workload, session, and runtime to evolve according to their own lifecycles. It also allows identity infrastructure to support downstream accountability, delegation, and governance without pulling those higher-layer responsibilities back into the identity system itself.

\section{Conclusion}\label{conclusion}

As some AI systems move from transient model invocations toward long-lived operation and sustained digital activity, identity infrastructure must do more than verify whether a particular Credential or protocol request is valid. It must also determine \textbf{which persistent subject, behind changing Credentials, Clients, sessions, and runtimes, is actually being authenticated and attributed.}

SoulAuth answers this problem with \textbf{Actor-native Identity}. It treats \texttt{ActorIdentity} as the canonical continuity boundary inside the SoulAuth identity domain, allowing both Humans and long-lived AIActors to exist as first-class identity subjects while keeping HumanAccount, Credential, Client, IdentityBinding, and AuthSession separate in responsibility and lifecycle. Credential rotation, Client changes, session renewal, and runtime migration can occur without requiring the creation of a new Actor.

Around this model, SoulAuth organizes the Identity Domain, Authentication Core, AuthSession, tokens and federation, control, security, audit, and persistence as explicit logical responsibilities, and realizes the core design in an open-source Rust reference implementation. Architecture Conformance further maps selected invariants to runtime behavior, Schemas, machine-readable contracts, and verification evidence so that architectural claims can be traced into concrete software artifacts.

Methodologically, SoulAuth demonstrates a complete \textbf{Philosophical Engineering} path: conceptual clarification identifies the structure of the persistent identity subject; that normative judgment is translated into system boundaries; engineering implementation, conformance testing, and fixed artifacts then make it possible to inspect whether the implementation remains faithful to the original definition. For Human-AI collaboration, this turns the question of ``who is the persistent subject?'' into an infrastructure fact that can be deployed, authenticated, audited, and verified, providing a more stable identity basis for downstream authorization, delegation, accountability, and governance.

Evaluation of the fixed \texttt{v0.1.0} source revision also shows that this translation is not yet complete. Several core Actor-native boundaries are realized in the implementation, while unified Credential modeling and historical attribution anchored to ActorIdentity remain only partially realized. We therefore report v0.1.0 as a \textbf{substantial reference implementation with partial architectural conformance}, not full conformance. This result demonstrates the practical value of Architecture Conformance: even when the system functions correctly, it can still reveal semantic gaps between the normative architecture and engineering reality.

SoulAuth's responsibility ends at identity and authentication. It does not define AI consciousness, legal personhood, or Mind Continuity, and it does not itself create delegation, governance, or execution Authority. Existing OAuth/OIDC, machine identity, Workload Identity, and delegation mechanisms can continue to perform their established roles, while Actor-native Identity provides a clear subject boundary for systems that require long-lived subject continuity and independent attribution.

From a security perspective, the purpose of this subject boundary is not to grant AIActors additional powers. It is to ensure that cryptographic proofs, Authentication Results, sessions, protocol projections, and historical records remain attributable to the correct persistent subject.

For the broader problem of Humans and long-lived AIActors entering a shared digital society, SoulAuth attempts to establish one foundational precondition: \textbf{before responsibility, authorization, and governance can be discussed, the system must be able to answer reliably ``who is being authenticated, and to whom should past authentication facts be attributed?''}

The core of the paper is therefore not the claim that ``AI also needs identity,'' but a more precise architectural judgment: \textbf{the canonical continuity boundary should be ActorIdentity, not Account, Credential, Client, Workload, AuthSession, Protocol Subject, or runtime instance.} Only after this boundary is stable can Credentials, protocols, runtimes, delegation, governance, and other higher-layer mechanisms build reliable relationships around the same persistent subject.

\begin{quote}
\textbf{SoulAuth can ultimately be reduced to one simple design principle: first determine who the Actor is, then determine how that Actor proves itself.}
\end{quote}

\section*{Artifact Availability}
\addcontentsline{toc}{section}{Artifact Availability}\label{artifact-availability}

The fixed public artifact for SoulAuth is:

\begin{itemize}
\tightlist
\item
  \textbf{Repository}: \url{https://github.com/TrantorLabs/SoulAuth}
\item
  \textbf{Release}: \texttt{v0.1.0}
\item
  \textbf{Release Date}: 2026-09-09
\item
  \textbf{Exact Git Commit}: \texttt{\seqsplit{aaad1abc52c8ab53c2a911f92e2698df95b0a17b}}
\item
  \textbf{Release Archive}: \texttt{soulauth-0.1.0.tar.gz}
\item
  \textbf{Archive SHA-256}: \texttt{\seqsplit{a674a16a8ffac3fcb515f61c55ada1a943e84cfc289423885435f7a6b85a9642}}
\item
  \textbf{License}: Apache-2.0
\item
  \textbf{GitHub Actions Evidence}: CI Run \texttt{\#34321518456}
\item
  \textbf{CI URL}: \texttt{https://github.com/TrantorLabs/SoulAuth/actions/runs/34321518456}
\end{itemize}

All evaluation results in Section 7 are bound to this fixed source revision. Complete commands, test identifiers, execution logs, and item-level engineering diagnostics for build, unit, integration, Architecture Conformance, and deployment checks are retained in the public repository and corresponding CI record rather than repeated in the paper. The artifact metadata provides a public, traceable, and repeatable evidence entry point; it is not presented as third-party independent artifact certification.

The paper's evidence relation can therefore be summarized as:

\begin{quote}
\textbf{Paper Claim → Fixed Release / Commit → Implementation \& Conformance Evidence → Observed Result → Disclosed Limitation.}
\end{quote}


\begin{thebibliography}{99}
\bibitem{ref1} B. Lampson, M. Abadi, M. Burrows, and E. Wobber. ``Authentication in Distributed Systems: Theory and Practice.'' \emph{ACM Transactions on Computer Systems}, 10(4):265--310, 1992.

\bibitem{ref2} N. Sakimura, J. Bradley, M. Jones, B. de Medeiros, and C. Mortimore. \emph{OpenID Connect Core 1.0 incorporating errata set 2}. OpenID Foundation, Final Specification, 15 December 2023.

\bibitem{ref3} SPIFFE Project. \emph{Secure Production Identity Framework for Everyone (SPIFFE)}. Cloud Native Computing Foundation, official specification, including SPIFFE ID, SVID, and Workload API specifications.

\bibitem{ref4} J. Salowey, Y. Rosomakho, and H. Tschofenig. \emph{Workload Identity in a Multi System Environment (WIMSE) Architecture}. IETF Internet-Draft draft-ietf-wimse-arch-08, 6 July 2026. Work in Progress.

\bibitem{ref5} Y. Ni and C. P. Liu. \emph{WIMSE Applicability for AI Agents}. Individual Internet-Draft draft-ni-wimse-ai-agent-identity-02, 28 February 2026. Work in Progress; expired 1 September 2026.

\bibitem{ref6} M. Sporny, A. Guy, M. Sabadello, and D. Reed, eds.~\emph{Decentralized Identifiers (DIDs) v1.0: Core architecture, data model, and representations}. W3C Recommendation, 19 July 2022.

\bibitem{ref7} S. Rodriguez Garzon, A. Vaziry, E. M. Kuzu, D. E. Gehrmann, B. Varkan, A. Gaballa, and A. Küpper. ``AI Agents with Decentralized Identifiers and Verifiable Credentials.'' \emph{Proceedings of the 18th International Conference on Agents and Artificial Intelligence (ICAART 2026)}, vol.~1, pp.~252--259, 2026. doi:10.5220/0014234400004052. Also available as arXiv:2511.02841.

\bibitem{ref8} M. Xu, X. Liu, Y. Guo, C. Liu, Y. Zhang, and X. Cheng. ``AgentDID: Trustless Identity Authentication for AI Agents.'' arXiv:2604.25189, 2026. Accepted by ICDCS 2026.

\bibitem{ref9} R. R. Rodriguez Jr.~``Agent Identity URI Scheme: Topology-Independent Naming and Capability-Based Discovery for Multi-Agent Systems.'' arXiv:2601.14567, 2026.

\bibitem{ref10} Z. Zhao and R. Zhao. ``Runtime-Independent Persistent Agents: Preserving Identity, Memory, and Code Across Models, Harnesses, and Servers.'' arXiv:2609.00546, 2026.

\bibitem{ref11} T. South et al.~\emph{Identity Management for Agentic AI: The new frontier of authorization, authentication, and security for an AI agent world}. OpenID Foundation Whitepaper; arXiv:2510.25819, 2025.

\bibitem{ref12} T. South, S. Marro, T. Hardjono, R. Mahari, C. D. Whitney, D. Greenwood, A. Chan, and A. Pentland. ``Authenticated Delegation and Authorized AI Agents.'' arXiv:2501.09674, 2025.

\bibitem{ref13} S. Prakash. ``AIP: Agent Identity Protocol for Verifiable Delegation Across MCP and A2A.'' arXiv:2603.24775, 2026.

\bibitem{ref14} K. Huang, V. S. Narajala, I. Habler, and A. Sheriff. ``Agent Name Service (ANS): A Universal Directory for Secure AI Agent Discovery and Interoperability.'' arXiv:2505.10609, 2025.

\bibitem{ref15} T. Otsuka, K. Toyoda, and A. Leung. ``AI Identity: Standards, Gaps, and Research Directions for AI Agents.'' arXiv:2604.23280, 2026.

\bibitem{ref16} Y. A. Arbel, P. N. Salib, and S. Goldstein. ``How to Count AIs: Individuation and Liability for AI Agents.'' \emph{Boston College Law Review}, forthcoming; arXiv:2603.10028, 2026.

\bibitem{ref17} Y. He, Z. Shan, L. Luo, and W. Wang. ``Accountable yet Anonymous AI Agents - Split-Knowledge Binding in National Agent-Identity Layer in China.'' arXiv:2607.23207, 2026.

\end{thebibliography}
\end{document}